\documentclass[prd,twocolumn,superscriptaddress,floatfix,amsmath,amssymb,amsfonts,longbibliography,nofootinbib]{revtex4-2}

\usepackage{float} 

\usepackage{comment}
\usepackage[normalem]{ulem}
\usepackage[english]{babel}
\usepackage{graphicx}
\usepackage{dcolumn}
\usepackage{bm}
\usepackage{blindtext}
\usepackage{verbatim}
\usepackage{mathrsfs}
\usepackage{musicography}
\usepackage{amsmath}
\usepackage{dsfont}
\usepackage{cancel}
\usepackage{physics}
\usepackage{epstopdf}
\usepackage{mathtools}
\usepackage{color}
\usepackage[usenames,dvipsnames]{pstricks}
\usepackage{epsfig}
\usepackage{pst-grad} 
\usepackage{pst-plot} 
\usepackage{hyperref}
\usepackage{verbatim}
\usepackage{afterpage}
\usepackage{sidecap}

\hypersetup{
    colorlinks=true,
    linkcolor=blue,
    filecolor=red,      
    urlcolor=cyan,
}

\newcommand{\ii}{\mathrm{i}}

\newcommand{\tinyspace}{\mspace{1mu}}
\newcommand{\bignorm}[1]{\bigl\lvert\!\bigl\lvert\tinyspace #1 \tinyspace\bigr\rvert\!\bigr\rvert}

\begin{document}

\title{Entanglement structure of quantum fields through local probes}

\author{Bruno de S. L. Torres}
\email{bdesouzaleaotorres@perimeterinstitute.ca}\affiliation{Perimeter Institute for Theoretical Physics, Waterloo, Ontario, N2L 2Y5, Canada}
\affiliation{Department of Physics and Astronomy, University of Waterloo, Waterloo, ON N2L 3G1, Canada}
\affiliation{Institute for Quantum Computing, University of Waterloo, Waterloo, Ontario, N2L 3G1, Canada}

\author{Kelly Wurtz}
\email{kwurtz@uwaterloo.ca}
\affiliation{Perimeter Institute for Theoretical Physics, Waterloo, Ontario, N2L 2Y5, Canada}
\affiliation{Institute for Quantum Computing, University of Waterloo, Waterloo, Ontario, N2L 3G1, Canada}
\affiliation{Department of Applied Mathematics, University of Waterloo, Waterloo, Ontario, N2L 3G1, Canada}

\author{Jos\'{e} Polo-G\'{o}mez}
\email{jpologomez@uwaterloo.ca}
\affiliation{Perimeter Institute for Theoretical Physics, Waterloo, Ontario, N2L 2Y5, Canada}
\affiliation{Institute for Quantum Computing, University of Waterloo, Waterloo, Ontario, N2L 3G1, Canada}
\affiliation{Department of Applied Mathematics, University of Waterloo, Waterloo, Ontario, N2L 3G1, Canada}

\author{Eduardo Mart\'{i}n-Mart\'{i}nez}
\email{emartinmartinez@uwaterloo.ca}
\affiliation{Perimeter Institute for Theoretical Physics, Waterloo, Ontario, N2L 2Y5, Canada}
\affiliation{Institute for Quantum Computing, University of Waterloo, Waterloo, Ontario, N2L 3G1, Canada}
\affiliation{Department of Applied Mathematics, University of Waterloo, Waterloo, Ontario, N2L 3G1, Canada}

\begin{abstract}

We present a framework to study the  entanglement structure of a quantum field theory inspired by the formalism of particle detectors in relativistic quantum information. This framework can in principle be used to faithfully capture entanglement in a QFT between arbitrary-shaped regions of spacetime without encountering UV divergences, bypassing many of the issues typically present in other approaches. Our results also establish the limits of the efficiency of entanglement harvesting, and may also be used to motivate an operational definition of entanglement between spacetime subregions in field theory.

\end{abstract}

\maketitle

\section{Introduction}

The study of entanglement in relativistic quantum field theories (QFTs) plays an important role both in the very foundational aspects of QFT itself, as well as in our current explorations of the interface between quantum theory and gravity. On the field theory side, the existence of entanglement in the ground states of QFTs~\cite{ReehSchlieder, WernerVacuum, WittenEntanglement} underlies much of our understanding of how to make relativity and quantum theory consistent. Ground state entanglement is at the centre of phenomena ranging from the black-hole information loss problem~\cite{Preskill1992, HawkingLoss, BlackHoles1, BlackHoles2, BlackHoles3} or violations of energy conditions in quantum field theory in curved spacetimes~\cite{Prihadi_2022, Balakrishnan2019}, to applications in relativistic quantum information theory (e.g.,  extraction of quantum resources~\cite{EntanglementFarming}, noise assisted communication~\cite{Koji2020}, and quantum energy teleportation~\cite{HottaTeleportation, HottaTeleportationReview}, among many others). On the quantum gravity side, in the context of the AdS/CFT correspondence, entanglement has been proposed as one key component for the emergence of a smooth classical spacetime geometry from quantum degrees of freedom~\cite{RaamsdonkEntanglement, Faulkner2014, Lashkari2014}. 

Quantifying the entanglement present in a quantum field between different spacetime regions, however, is plagued with challenges. To begin with, even in the simple case of complementary regions (the only case in which entanglement entropy is a valid entanglement measure), the entanglement entropy across the bipartition is UV-divergent. Further, the regularization of the entanglement entropy quickly becomes intractable for arbitrary region shapes or interacting theories. The problem worsens if one wants to quantify entanglement between two non-complementary regions, since their joint state is not pure and the so-called `entanglement entropy' stops being a valid entanglement measure. In those cases, other quantities---such as the mutual information or the relative entropy of entanglement---are often used to quantify the correlations between the regions~\cite{Tonni2011, Hayden2013, Fonda2015, HollandsRelativeEntropyEntanglement}. However, the former is not an entanglement measure in general, and the latter is exceedingly hard to compute in practice, with only upper and lower bounds being known in a few special cases.

In Relativistic Quantum Information, the protocol of entanglement harvesting was developed to access the entanglement contained in quantum fields by probing them with \emph{particle detectors}. Particle detectors are localized probes which couple to the field in finite regions of space and time~\cite{Unruh1976,DeWitt,Langlois}. Through coupling with a field, a pair of particle detectors can become entangled even when they are spacelike separated~\cite{Valentini1991,Reznik2003,Reznik2,Pozas-Kerstjens:2015}. Since the detectors interact with the field locally \cite{TalesBruno2, PipoMaria}, the entanglement picked up by the spacelike separated detectors can only be due to pre-existing entanglement in the initial state of the field. Crucially, the formalism of particle detectors allows us to easily handle scenarios of entanglement harvesting in which the detectors couple to arbitrarily shaped regions even in non-stationary regimes. What is more, the coupling of particle detectors is, by construction, devoid of the UV-divergences that other analyses  
suffer. Moreover, particle detector models are not just a theoretical construct: they have been proven to be good models for the kind of sensors that, in practice, are used to measure quantum fields in the lab, such as atomic probes~\cite{eduardoOld,Pablo,Richard}.

While it may be tempting to interpret the entanglement harvested by detectors as a direct witness of the entanglement structure of the QFT, there are some important obstacles to making this direct connection. First, detectors suffer from local vacuum noise that hinders their ability to efficiently extract field entanglement~\cite{Ruep_2021, DanBruno}. Second, detectors have a finite number of degrees of freedom, whereas the entanglement in quantum fields is distributed among its infinite-dimensional phase space. As a consequence, the detectors' specific coupling may condition what degrees of freedom in the field they are actually extracting entanglement from. Finally, even if we overlooked these issues, we have no a priori way to know if the detectors' entanglement is in any way faithfully capturing all the entanglement present in the field regions.

In this paper we analyze how to optimally transfer features of the entanglement structure of a quantum field to particle detectors that couple to the field in localized regions of spacetime. We will provide upper bounds to how much entanglement a fixed number of detectors can extract from the field in a given region. We will also show how the spatial distribution of the extracted entanglement behaves and we will prescribe the coupling between detectors and field (for any local subregion) which will best reflect the entanglement that the given spatial region had with its complement.

The paper is organized as follows. Section~\ref{DetectorSection} reviews the basic elements of a measurement framework for quantum fields using particle detectors, and lays out the setup for the detector and field that we will employ in this paper. Section~\ref{DiscreteFieldSection} describes a discretized version of the field theory as a chain of harmonic oscillators, which will be convenient for the use of numerical methods in the description of the system. We also review basic features of the entanglement between subregions in the harmonic chain which we aim to capture with the aid of particle detectors. Section~\ref{OptimalCouplingSection} establishes the bound on how much entanglement can be obtained from a given region of the field, given the number of degrees of freedom accessible in the probes. This will lead to a prescription for the coupling between the field and a given number of detectors, in order for the final state of the probes to reflect as much as it can of the entanglement between the coupling region and the rest of space. As a proof of principle, we show in Section~\ref{AreaLawSection} that, with the preferred choice of coupling outlined in Section~\ref{OptimalCouplingSection}, the entanglement between the detector degrees of freedom and the rest of space displays an area-law behavior, as expected for entanglement entropies between complementary regions in ground states of systems with short-range interactions. Section~\ref{sectioncontinuum} addresses how the results obtained in Sections~\ref{OptimalCouplingSection} and~\ref{AreaLawSection} can be adapted to reflect the case of a continuum field theory, even though they were mostly explored in the context of a discretized version of the field as described in Sec.~\ref{DiscreteFieldSection}. We end in Section~\ref{ConclusionSection} with our concluding remarks, including some comments on future directions. 

Appendices~\ref{GQMSection} and~\ref{AppendixProof} provide a basic overview of the important concepts and technical results in the Gaussian formalism of bosonic quantum systems, which we make heavy use of throughout the manuscript. We work in natural units where $\hbar = c = 1$.

\section{Measuring quantum fields with particle detectors}\label{DetectorSection}

Modelling the action of measurements on relativistic quantum fields poses additional problems on top of the foundational issues already present in the measurement problem of non-relativistic quantum mechanics. To begin with, even as a matter of principle, one cannot assume that finite-rank projective measurements (PVMs) of localized field observables exist. In fact, it is well known that they do not exist even for the simplest of field theories, since their existence is fundamentally incompatible with the relativistic nature of the theory, and there is no known cure for this~\cite{ReehSchlieder,Schlieder1968,Hellwig1970formal,Redhead1995}. It has been argued that particle detector models provide a tool to model the measurement of quantum fields\footnote{Note that there are other approaches to the measurement problem in QFT besides the particle detector approach. Another promising avenue is the Fewster-Verch framework. For more information see~\cite{MariaDoreen,fewster1,fewster2}.} that alleviates this kind of problem~\cite{PipoMaria, JoseFramework} and that directly connects with the way in which quantum fields are actually measured in a lab~\cite{JoseFramework}. 

In this section we will introduce particle detector models, explain how they interact with quantum fields, and set up the stage for the formalism that we will employ to study the entanglement structure of quantum fields through particle detectors. 

A particle detector is a localized nonrelativistic quantum system which couples to a relativistic quantum field in a localized region of spacetime. In the most basic setting, the detector follows a classical trajectory given by some timelike path in spacetime. The detector has internal degrees of freedom whose dynamics can be prescribed by a free Hamiltonian generating translations with respect to the detector's proper time $\tau$. For detectors in arbitrary trajectories in possibly curved spacetimes, the coupling between detector and field can be described by an interaction Hamiltonian which, in a given coordinate system $\mathsf{x}=(t, \bm{x})$ in a $D=n+1$ dimensional spacetime, takes the form
\begin{equation}\label{generalintH}
    \hat{H}_{\text{I}}(t) = \lambda \sum_i\int_{\Sigma_t}\!\!\! \dd^n \bm{x} \,\sqrt{-g(\mathsf{x})}\, \Lambda_i(\mathsf{x})\hat{\mu}_i[\tau(\mathsf x)]\hat{\mathcal{O}}_i(\mathsf{x}).
\end{equation}
Here, $g$ is the determinant of the metric in the coordinates $(t, \bm{x})$, and $\Sigma_t$ is a codimension-$1$ spacelike surface fixed by a constant value of the timelike coordinate $t$. Each $\hat{\mu}_i(\tau)$ is some operator acting on the Hilbert space of the detector, $\tau$ is the proper time of the detector, and $\hat{\mathcal{O}}_i(\mathsf{x})$ is some field observable being probed. $\lambda$ is an overall coupling constant, and $\Lambda_i(\mathsf{x})$ is a spacetime smearing function which strongly localizes the coupling between the detector and field observable $\mathcal{O}_i(\mathsf{x})$ to the vicinity of the detector's trajectory. One can write this in more succinct form as
\begin{equation}
    \hat{H}_{\text{I}}(t) =\int_{\Sigma_t} \dd^n \bm{x}\sqrt{-g}\, \hat{h}_{\text{I}}(\mathsf{x}),
\end{equation}
where $\hat{h}_{\text{I}}(\mathsf{x})\coloneqq \lambda\sum_i\Lambda_i(\mathsf{x})\hat{\mu}_i[\tau(\mathsf{x})]\hat{\mathcal{O}}_i(\mathsf{x})$ is an interaction Hamiltonian weight\footnote{Note that the (scalar) Hamiltonian weight $\hat h$ is related to the (pseudoscalar) Hamiltonian density $\hat{\mathfrak{h}}$ by $\hat{\mathfrak{h}}=\sqrt{-g}\,\hat{h}$, see~\cite{TalesBruno1} for more details.}. Once the interaction is given, the dynamics can be described in the interaction picture by the time evolution operator
\begin{align}
    \hat{U} &= \mathcal{T}\exp\left(-\ii\int \dd t\hat{H}_\text{I}(t)\right) = \mathcal{T}\exp\left(-\ii\int \dd V\hat{h}_\text{I}(\mathsf{x})\right),
\end{align}
where $\dd V= \dd^{D}\mathsf{x} \,\sqrt{-g}$ is the invariant spacetime volume element, and $\mathcal{T}\exp$ denotes the time ordered exponential. 

The initial state of the full system is taken to be a completely uncorrelated state of detector and field,
\begin{equation}
    \hat{\rho}_0 = \hat{\rho}_{\text{d}, 0}\otimes \hat{\rho}_\phi,
\end{equation}
where, commonly, $\hat{\rho}_{\text{d},0}$ is chosen to be the ground state of the detector's internal Hamiltonian, and $\hat{\rho}_\phi$ the field's vacuum state. After the interaction takes place, the final state of the detector is
\begin{equation}\label{detectorfinalstate}
    \hat{\rho}_{\text{d}} = \Tr_{\phi}\left[\hat{U}\left(\hat{\rho}_{\text{d}, 0}\otimes \hat{\rho}_\phi\right)\hat{U}^\dagger\right],
\end{equation}
which is often treated perturbatively by writing $\hat{U}$ in terms of a Dyson expansion, and collecting each power of the coupling constant $\lambda$. 

In general, the final state of the detector in Eq.~\eqref{detectorfinalstate} will depend nontrivially on details of the field's initial state $\hat{\rho}_\phi$. By engineering a suitable spacetime smearing function for the coupling, and then analyzing the final state of the detector, it is possible to indirectly infer properties of the field being probed~\cite{DanIreneML}. 

In this paper, we will focus on probing a free real scalar field in Minkowski space, whose free dynamics in $n+1$ spacetime dimensions can be described in a given inertial coordinate system $(t, \bm{x})$ by the Hamiltonian
\begin{equation}\label{fieldhamiltonian}
    \hat{H}_{\phi} = \dfrac{1}{2}\int \dd^n \bm{x}\left(\hat{\pi}^2 + (\nabla \hat{\phi})^2 + m^2\hat{\phi}^2\right)
\end{equation}
where $m$ is the field mass and $\hat{\pi} \coloneqq \partial_t \hat{\phi}$ is the field's canonical momentum. $\hat{\phi}$ and $\hat{\pi}$ satisfy the equal-time canonical commutation relations 
\begin{equation}
    \comm{\hat{\phi}(t, \bm{x})}{\hat{\pi}(t, \bm{x}')} = \ii \delta^{(n)}(\bm{x}-\bm{x}').
\end{equation}
The local probes (i.e., the detectors) will be taken to be harmonic oscillators. For simplicity, we will consider them to be comoving with the frame $(t,\bm x)$. For each detector we have a pair of dimensionless quadratures $(\hat{q}_\text{d}, \hat{p}_\text{d})$ satisfying $[\hat{q}_\text{d}, \hat{p}_\text{d}]=\ii \openone$ and free detector Hamiltonian
\begin{equation}\label{freedetectorH}
    \hat{H}_\text{d} = \dfrac{\Omega}{2}\left(\hat{p}^2_\text{d} + \hat{q}^2_\text{d}\right),
\end{equation}
which will couple to the field through an interaction that we will generally consider to be of the form
\begin{align}\label{fieldcoupling}
    \hat{H}_\text{I} = \lambda\chi(t)&\bigg(\lambda_q\hat{q}_\text{d}\int\dd^n \bm{x}\, f(\bm{x})\hat{\pi}(t, \bm{x}) \\
    &\;\;- \lambda_p\hat{p}_\text{d}\int\dd^n \bm{x}\, g(\bm{x})\hat{\phi}(t, \bm{x})\bigg).\nonumber
\end{align}
Here, $\lambda$ is a dimensionless coupling constant, $\lambda_q$ and $\lambda_p$ are scales adjusting the dimensions of each term\footnote{Since we consider in this paper that the switching and smearing functions are $L^1$-normalized functions, \mbox{$[\chi(t)]=L^{-1}$}, \mbox{$[f(\bm x)]=[g(\bm x)]=L^{-n}$}, this means that \mbox{$[\lambda_q]=[\pi]^{-1}=L^{\frac{n+1}{2}}$} and \mbox{$[\lambda_p]=[\phi]^{-1}=L^{\frac{n-1}{2}}$}. Some natural parameters that could be used to adjust the units would be the field mass, $m$, or the detector's frequency, $\Omega$, with $[m]=[\Omega]=L^{-1}$.}, and $\chi(t)$ is a switching function which dictates the time dependence of the coupling between the detector and the field. $f(\bm{x})$ and $g(\bm{x})$ correspond to the (spatial) smearing functions---the spatial profile of the interaction---for the coupling to the field and its conjugate momentum, respectively. Naturally, if there are multiple detectors, the free detector Hamiltonian contains a sum of terms like~\eqref{freedetectorH}, and the interaction Hamiltonian will contain a sum of terms like~\eqref{fieldcoupling}, with possibly different choices of $f$ and $g$ for each detector.

If the smearing functions $f(\bm{x})$ and $g(\bm{x})$ satisfy
\begin{equation}
    \lambda_q \lambda_p\int \dd^n \bm{x} \,f(\bm{x})g(\bm{x}) = 1,
\end{equation}
then we can interpret the operators
\begin{align}
    \hat{Q} &\coloneqq \lambda_p \int \dd^n \bm{x}\, g(\bm{x})\hat{\phi}(t, \bm{x}), \label{defnormalmodescontinuumQ}\\
    \hat{P} &\coloneqq \lambda_q \int \dd^n \bm{x}\, f(\bm{x}) \hat{\pi}(t, \bm{x}) \label{defnormalmodescontinuumP}
\end{align}
as the quadratures that define one single mode in the field, since they satisfy $[\hat{Q},\hat{P}] = \ii\mathds{1}$. With this, the interaction Hamiltonian~\eqref{fieldcoupling} can be seen as a coupling between the detector and one particular mode of the field.

The time evolution operator for the dynamics between detector and field in this setup simplifies drastically when we consider the interaction Hamiltonian to be very intensely peaked around a very short period of time, with the final state being examined immediately after the interaction takes place. Mathematically, this amounts to taking $\chi(t) = \delta(t)$ in~\eqref{fieldcoupling}, where the internal dynamics generated by the free Hamiltonians can be neglected, and the full unitary time evolution operator will simply be given by
\begin{equation}
    \hat{U} = \exp\left[-\ii\lambda\left(\hat{q}_\text{d}\hat{P} - \hat{p}_\text{d}\hat{Q}\right)\right].
\end{equation}
This will simplify even further if we take $\lambda=-\pi/2$, in which case the dynamics implemented by
\begin{equation}\label{gaussianswap}
    \hat{U}_{\text{SWAP}} = \exp\left[\ii\,\dfrac{\pi}{2}(\hat{q}_\text{d} \hat{P} - \hat{p}_\text{d}\hat{Q})\right]
\end{equation}
will act on the quadratures as
\begin{equation}
\begin{array}{rcl}
   \hat{q}_\text{d}&\mapsto& \hat{Q}\\
   \hat{Q}&\mapsto& -\hat{q}_\text{d}\\
   \hat{p}_\text{d}&\mapsto& \hat{P}\\
   \hat{P}&\mapsto& -\hat{p}_\text{d}
\end{array}
\end{equation}
and will therefore swap the phase spaces between the mode $(\hat{q}_\text{d}, \hat{p}_\text{d})$ and the mode $(\hat{Q}, \hat{P})$. This is why we refer to the Gaussian unitary~\eqref{gaussianswap} as a \emph{swap operator}. With the dynamics reduced to a swap between the detector and some mode of the field, we guarantee that the final state of the detector will be exactly the initial state of the field mode that is probed. In other words, the final state of the detector can be immediately identified as the reduced state of the field in the selected mode $(\hat{Q}, \hat{P})$. Therefore, the swap operation allows us to speak about what a particle detector can measure in terms of specific field modes. 

\section{Entanglement in the ground state of a QFT}\label{DiscreteFieldSection}

It will be convenient for our purposes to consider first a setup where we can use the tools of Gaussian quantum mechanics. In order to do that we replace the field theory (which contains infinitely many degrees of freedom) by a lattice of harmonic oscillators, equivalent to introducing an IR and a UV cutoff in the theory. We will address how the results generalize to the continuum case in Sec.~\ref{sectioncontinuum}. The Hamiltonian~\eqref{fieldhamiltonian} then takes the discrete form
\begin{equation}\label{discretefieldhamiltonian}
    \hat{H}_{\phi} = \dfrac{1}{2}\sum_i\omega\left(\hat{q}_i^2 + \hat{p}_i^2\right) - \dfrac{\alpha}{2}\sum_{\langle i,j\rangle}\hat{q}_i\hat{q}_j.
\end{equation}
The first term is a sum over all the sites of the \mbox{$n$-dimensional} lattice, and the notation $\langle i,j\rangle$ in the second term denotes a sum over nearest neighbours. Each pair $(\hat{q}_i, \hat{p}_i)$ corresponds to dimensionless quadratures, and $\omega$ represents the decoupled (free) frequency of each local oscillator, which is the same for all sites. One can see the lattice Hamiltonian~\eqref{discretefieldhamiltonian} as a discrete approximation\footnote{Note that \eqref{discretefieldhamiltonian} is not exactly the result of applying UV and IR cutoffs to \eqref{fieldhamiltonian}: there is an approximation in taking the derivative of the field to be given by a first neighbours discrete derivative. This is not a problem because the continuum limit of this approximation is still the continuous derivative and \eqref{discretefieldhamiltonian} indeed becomes \eqref{fieldhamiltonian} in the limit $\varepsilon\to 0$. For a deeper discussion on this subtlety see, e.g.,~\cite{Dan2022Part2}.} to the Hamiltonian~\eqref{fieldhamiltonian} through the replacements 
\begin{align}
    \text{spatial coordinate}\,\,\bm{x}&\mapsto \text{discrete label} \,\,\,i, \nonumber\\
    \int \dd^n x&\mapsto \varepsilon^n \sum_i, \nonumber \\
    \delta^{(n)}(\bm{x}-\bm{y})&\mapsto \dfrac{1}{\varepsilon^n}\delta_{ij},\\
    (\nabla\hat{\phi}(\bm{x}))^2&\mapsto \dfrac{1}{2\varepsilon^2}\sum_{j}\left(\hat{\phi}_i - \hat{\phi}_j\right)^2\nonumber
\end{align}
where $\varepsilon$ is the short-distance cutoff length scale, and in the last expression the sum in $j$ runs over all nearest neighbours of the site with label $i$. With these replacements, the free frequency $\omega$ is related to the field mass $m$ and the cutoff $\varepsilon$ by 
\begin{equation}\label{latticefrequency}
    \omega^2 = m^2 + \dfrac{2n}{\varepsilon^2}.
\end{equation}
The quadratures $(\hat{q}_i, \hat{p}_i)$ at each lattice site are related to the discretized version of $\hat{\phi}$ and $\hat{\pi}$ by
\begin{align}
    \hat{\phi}_i &= \dfrac{1}{\sqrt{\omega\varepsilon^n}}\hat{q}_i, \label{lattice position}\\
    \hat{\pi}_i &= \sqrt{\dfrac{\omega}{\varepsilon^n}}\hat{p}_i, \label{lattice momentum}
\end{align}
and the effective lattice coupling constant $\alpha$ is given by
\begin{equation}\label{latticecoupling}
\alpha = \dfrac{1}{\omega\varepsilon^2}.
\end{equation}
Since $m^2\geq 0$, Eqs.~\eqref{latticefrequency} and~\eqref{latticecoupling} imply that, for a given value of $\omega$, the coupling $\alpha$ is constrained by \mbox{$\alpha \leq \alpha_c \equiv \omega/2n$}, with the inequality being saturated in the case of a massless field. If $\alpha$ were greater than $\alpha_c$, the Hamiltonian~\eqref{discretefieldhamiltonian} would not be bounded from below.

Imposing a short-distance cutoff places a finite number of modes per unit volume in the system. By then adding an IR cutoff (say, by placing the field in a box with a large but finite side length) the total number of oscillators is made finite, making the problem more amenable to Gaussian methods. Since the interactions are short-range (only nearest neighbours coupling), the local features of the system will not be affected by the presence of the boundary for subregions of sufficiently small size compared to the full size of the system.

Since the Hamiltonian is quadratic, the ground state of the lattice is Gaussian and hence fully characterized by its covariance matrix, which can be expressed in closed form in terms of the parameters of the Hamiltonian~\eqref{discretefieldhamiltonian}. This is most easily done by writing $\hat{H}_\phi$ as
\begin{equation}
    \hat{H}_{\phi} = \dfrac{\omega}{2}\sum_i \hat{p}_i^2 + \dfrac{1}{2}\hat{\bm{q}}^\intercal\bm{V}\hat{\bm{q}},
\end{equation}
where the symmetric matrix $\bm{V}$ has its elements defined by
\begin{equation}
    V_{ij} = \begin{cases}
    \omega, \,\,\,\,\,i=j \\
    -\alpha, \,\,\,\,\,i, j \,\,\,\text{nearest neighbours}\\
    0 \,\,\,\,\,\,\text{otherwise.}
    \end{cases}
\end{equation}
Then, in the ordered basis $(\hat{\bm{q}}, \hat{\bm{p}})$ the ground state covariance matrix $\sigma_\phi$ can written as~\cite{Audenaert2002}
\begin{equation}\label{vacuumcovmatrix}
    \sigma_\phi = \begin{pmatrix}\sqrt{\omega \bm{V}^{-1}} & \mathbf{0} \\ \mathbf{0} & \sqrt{\omega^{-1}\bm{V}} \end{pmatrix}.
\end{equation}
Once the covariance matrix of the ground state is known, characterizing entanglement measures is in principle straightforward~\cite{Simon2000,Vidal2001,Adesso2005}. Since the ground state is pure, for any mode decomposition associated with some canonical coordinate system in phase space, the entanglement between a set of modes and its complement can be quantified by the entanglement entropy, which is the von Neumann entropy of either of the subsystems in the bipartition. The entanglement entropy is fully determined by the symplectic eigenvalues of the reduced covariance matrix of either subsystem in a Gaussian state. 

When we are interested in the entanglement between two sets of modes that together do \emph{not} form the whole system, we will adopt the logarithmic negativity, since their joint state will in general not be pure, and therefore the `entanglement entropy' would not be a measure of entanglement.  In contrast, the logarithmic negativity is a faithful, additive, entanglement monotone for bipartitions between a single mode and an arbitrary number of other modes for Gaussian states, whether they are pure or mixed~\cite{Plenio2005}. Even in the cases where negativity is not a faithful monotone, it is still a helpful quantity to characterize entanglement, since  non-zero negativity is a sufficient condition for nonseparability, and logarithmic negativity is directly quantifying the amount of distillable entanglement in a bipartition~\cite{Plenio2005, Audenaert2003}. Given that  bound entanglement in Gaussian states is rare~\cite{Werner2001}, for our purposes logarithmic negativity accurately quantifies entanglement. The logarithmic negativity for a given bipartition of a composite Gaussian quantum state can be easily computed from the symplectic spectrum of the partially transposed covariance matrix. We show the details of its derivation in Appx.~\ref{GQMSection} and its expression is given in Eq.~\eqref{lognegativitygaussian}.

One of the most fundamental observations about ground states of a variety of many-body systems with short-range interactions is the fact that the entanglement between complementary subregions, whether quantified by entanglement entropy or logarithmic negativity, obeys an area law---i.e., they both grow with the area of the surface separating the two regions~\cite{Plenio2005arealaw,Cramer2006,Eisert2010}. For a free field theory in the continuum, the entropy of a subregion is formally divergent due to the growing amount of entanglement between degrees of freedom at arbitrarily short distances across the boundary that separates the subregion from its complement. However, by placing a UV cutoff, one can show that the leading-order divergence to the entanglement entropy is indeed proportional to the area as the cutoff is sent to zero---as can be verified analytically, for instance, by Euclidean path integral calculations via the replica trick (see. eg.,~\cite{EntropyHolographyReview} for a review). We will come back to how this area-law behavior can be optimally transferred to detectors in Section~\ref{AreaLawSection}.

\section{Upper bound to entanglement extraction with particle detectors}\label{OptimalCouplingSection}

In this section, we will prescribe how to select, within any localized region $A$ of a harmonic lattice and for a given integer $N$, the set of $N$ modes supported in $A$ which are most entangled with the region's complement $\bar A$. This will naturally lead to a prescription for the set of modes that local probes with $N$ bosonic degrees of freedom must couple to in order to be able to extract the most entanglement with $\bar A$. As we will see in Section~\ref{sectioncontinuum}, since a preferred choice of modes can be reinterpreted as a preferred smearing function for the detector's coupling to the field, these results can be directly rephrased as a recipe for the structure of the coupling between probes and field that will optimally transfer entanglement from the field to the detectors.

The key to the recipe we are about to describe comes from the following observation, which derives from how the entanglement contained in a globally pure Gaussian state is distributed across (Williamson) normal modes\footnote{(Williamson) normal modes should not be confused with the mechanical normal modes of a quadratic interaction Hamiltonian (which are decoupled modes of free oscillation). In this manuscript, following common convention, we will use the name normal modes always referring to the modes obtained under a canonical change of variables for which the covariance matrix is diagonal (i.e. the modes are not correlated).} supported in complementary regions.

\textbf{Theorem}: Let $A$ be a local subregion containing $N_A$ modes of an overall pure Gaussian state, and let $\bar{A}$ be the region that is complementary to $A$. Then, the entanglement (as quantified by the logarithmic negativity) between the region $\bar{A}$ and any collection of $m\leq N_A$ modes supported entirely within $A$ is upper-bounded by the entanglement between $\bar{A}$ and the set of $m$ most mixed normal modes of the reduced covariance matrix of subregion $A$.

The proof goes as follows. First we note that, by definition, the covariance matrix of subregion $A$ in the basis of normal modes takes the form of a direct sum, which means that all modes in that basis are completely uncorrelated with each other. Because the joint state of $A\bar{A}$ is pure, however, the normal modes of $A$ must be purified by another set of modes that are fully supported on $\bar{A}$. In fact, the purification works mode-by-mode: for each normal mode of $A$, there is a separate normal mode of the complementary region $\bar{A}$ that shares all of its entanglement with the given mode in $A$, such that the two modes paired together are in a product state with the rest of the system\footnote{This is actually a special case of a more general fact: in a pure Gaussian state, for \emph{any} single mode (not necessarily a normal mode of a subregion's covariance matrix), one can define a second mode with the property that the joint state of the two modes fully decouples from the rest. This pair of modes that completely purify each other is known as a \emph{partner mode} pair~\cite{HottaPartnerMode1, HottaPartnerMode2, HottaPartnerMode3}. What is special about partner mode pairs comprising normal modes of complementary subregions is that, in this case in particular, the spatial supports of each mode in the pair do not overlap, and each mode is completely uncorrelated from the rest of the modes in the region where it is supported~\cite{Botero2002,Botero2004,Wolf2008}.}. 

In order to analyze entanglement between normal modes in $A$ and $\bar A$, we first build the partial transpose $\tilde{\sigma}_{A\bar{A}}$ of the covariance matrix ${\sigma}_{A\bar{A}}$ (with respect to  the modes in the region $\bar{A}$) in the basis of normal modes of $A$ and $\bar{A}$. In this basis, it is particularly easy to see that the symplectic spectrum of $\tilde{\sigma}_{A\bar{A}}$---from which the logarithmic negativity between $A$ and $\bar{A}$ is computed---can be fully expressed in terms of the symplectic eigenvalues of $\sigma_A$. More specifically, the $i$-th lowest symplectic eigenvalue $\tilde{\nu}_i$ of $\tilde{\sigma}_{A\bar{A}}$ is related to the $i$-th \emph{largest} symplectic eigenvalue $\nu_i$ of $\sigma_A$ through 
\begin{equation}\label{CitationNeeded}
    \Tilde{\nu}_i = \nu_i - \sqrt{\nu_i^2 - 1}.
\end{equation}
A proof of this statement can be found in Appendix~\ref{AppendixProof}. One readily sees that $\nu_i\geq 1$ automatically guarantees that $\Tilde{\nu}_i \leq 1$, and that $\Tilde{\nu}_i$ is a monotonically decreasing function of $\nu_i$. 

A change of basis on the modes supported in $A$ will transform the covariance matrix of $A\bar{A}$ as
\begin{equation}
    \sigma_{A\bar{A}}' = \left(S_A\oplus \mathds{1}_{2N_{\bar{A}}}\right)\sigma_{A\bar{A}}\left(S_A^\intercal\oplus \mathds{1}_{2N_{\bar{A}}}\right),
\end{equation}
where $S_A$ is some symplectic transformation acting only on $A$. The logarithmic negativity between regions $A$ and $\bar{A}$ can also be computed from $\sigma'_{A\bar{A}}$ according to the symplectic spectrum of $\Tilde{\sigma}'_{A\bar{A}}$, obtained by
\begin{align}
    \Tilde{\sigma}'_{A\bar{A}} &= \left(\mathds{1}_{2N_A}\oplus T_{\bar{A}}\right)\sigma'_{A\bar{A}}\left(\mathds{1}_{2N_A}\oplus T_{\bar{A}}\right)\nonumber \\
    &= \left(S_A\oplus \mathds{1}_{2N_{\bar{A}}}\right)\Tilde{\sigma}_{A\bar{A}}\left(S^\intercal_A\oplus \mathds{1}_{2N_{\bar{A}}}\right).
\end{align}
Here, $T_{\bar{A}}$ is the operator that performs a local time reversal\footnote{Local time reversal is the phase-space version of partial transposition on the Hilbert space, in terms of which we define the logarithmic negativity in general. For more details, see Appx.~\ref{GQMSection}.} on the modes in $\bar{A}$---i.e., it sends $\hat{P}$ to $-\hat{P}$ for every momentum operator supported in region $\bar{A}$, and does nothing to the position operators. In the ordered basis $(\hat{Q}^i, \hat{P}_i)$ for the subregion $\bar{A}$, it corresponds to the matrix
\begin{equation}
    T_{\bar{A}} = \bigoplus_{i = 1}^{N_{\bar{A}}} \begin{pmatrix}1 & 0 \\ 0 & -1 \end{pmatrix}.
\end{equation}
Since symplectic transformations do not change the symplectic spectrum, we know that the symplectic eigenvalues of $\Tilde{\sigma}'_{A\bar{A}}$ are the same as those of $\Tilde{\sigma}_{A\bar{A}}$, as they should be.

\begin{SCfigure*}[0.25][h]
\begin{wide}
    \hspace{-.7cm}
    \includegraphics[width=0.78\textwidth]{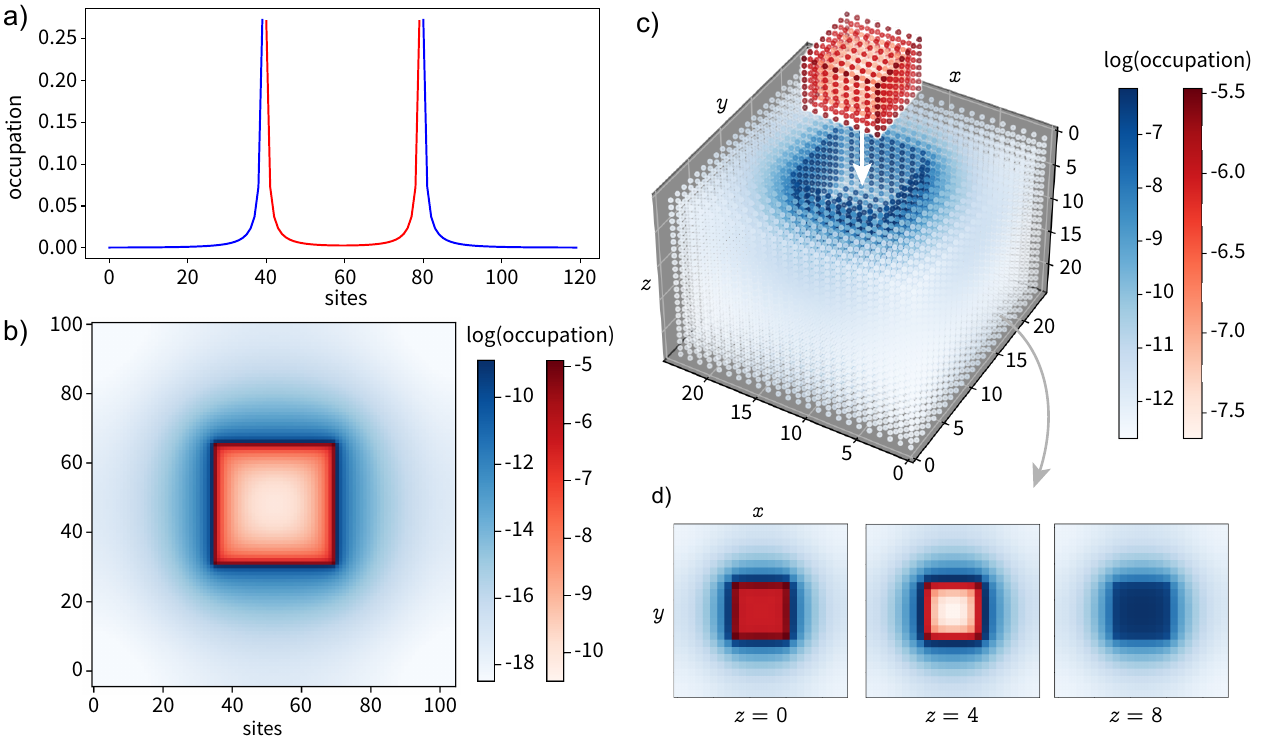}
    \caption{Spatial profile of the most mixed normal mode of a given region (in red) and of its complement (in blue) as given by the occupation function $F_i$. Subfigures a) and b) show this for a lattice in 1 and 2 spatial dimensions respectively. Subfigures c) and d) are both in 3D, with c) being a full three-dimensional visualization of the mode and its complement, and d) showing two-dimensional slices at different heights of the plot in the top. }
    \label{fig:occupationPlots}
\end{wide}
\end{SCfigure*}

Now consider any collection of $m$ modes supported within region $A$. These can be defined by applying an appropriate symplectic transformation that sends the basis of normal modes of subsystem $A$ to a basis that contains the list of desired modes, after which we discard the extra $N_A-m$ degrees of freedom. The logarithmic negativity between the chosen collection of modes in $A$ and the entire complementary region $\bar{A}$ is similarly extracted from the symplectic spectrum of the covariance matrix obtained after discarding the rows and columns of $\Tilde{\sigma}'_{A\bar{A}}$ that correspond to the modes we threw away. 

We can now invoke the interlacing theorem for symplectic eigenvalues. This theorem (which we state in more detail in Appx.~\ref{AppendixProof}) guarantees that the $j$-th lowest symplectic eigenvalue of $\tilde{\sigma}'_{A\bar{A}}$ lower-bounds the $j$-th lowest symplectic eigenvalue of any matrix obtained from $\tilde{\sigma}'_{A\bar{A}}$ by discarding the rows and columns associated with one of its modes. By applying this $N_A - m$ times, and using that the logarithmic negativity for a Gaussian state is given by Eq.~\eqref{lognegativitygaussian} with $F(x)$ being a monotonically decreasing function in the interval $(0, 1)$, it then follows that the maximum logarithmic negativity between $m$ modes of region $A$ and the complementary region $\bar{A}$ cannot be larger than the logarithmic negativity between $\bar{A}$ and the $m$ most mixed normal modes of $A$. This completes the proof that, indeed, the largest amount of entanglement between a collection of $m$ modes in $A$ and the complementary region $\bar{A}$ is attained by choosing the $m$ most mixed normal modes of region $A$.

The quadratures for the normal modes $(\hat{Q}_i, \hat{P}_i)$ can be expressed as linear combinations of the quadratures of the local harmonic oscillators that build the subregion $A$ being probed. That is, for any normal mode $(\hat{Q}, \hat{P})$, we can write
\begin{equation}\label{defnormalmodes}
    \hat{P} = \sum_i f_i \hat{p}_i, \qquad
    \hat{Q} = \sum_i g_i \hat{q}_i
\end{equation}
where $f_i$ and $g_i$ are the spatial profiles of the normal mode in momentum and position, respectively.
The spatial profile of the most mixed normal mode of a typical subregion in $1$, $2$, and $3$ spatial dimensions are shown in Fig.~\ref{fig:occupationPlots}. In each case, the parameters $\omega$ and $\alpha$ from Eq.~\eqref{discretefieldhamiltonian} were chosen such that the ratio $\alpha/\omega$ is within $0.1\%$ of the critical value $1/2n$, with $n$ being the number of spatial dimensions of the lattice. This corresponds to the (almost) massless limit of the field theory\footnote{Notice that if we increased the mass of the field, since the correlations of the region with its complement decay exponentially with the field mass, we would expect the most mixed modes to quickly become more localized around the boundary.}. The quantity being plotted in Fig.~\ref{fig:occupationPlots} is the \emph{mode occupation function}, which is given at each site $i$ by $F_i \coloneqq f_ig_i$, with $g_i$ and $f_i$ defined as in~\eqref{defnormalmodes}. It is clear that the support of the normal modes is strongly concentrated near the boundary of the region. This is to be expected, since the Hamiltonian only contains nearest-neighbour interactions, and therefore the entanglement present in the ground state between a region and its complement is mostly shared across the region's boundary.

Once we have established the list of $m$ modes within region $A$ that are most entangled with $\bar{A}$, it is then straightforward to prescribe an idealized coupling that transfers their entanglement to a series of $m$ probes whose coupling to the lattice is restricted to lie strictly on $A$. For a detector with $m$ bosonic degrees of freedom---or, equivalently, for $m$ single-mode detectors---given by $(\hat{q}_i, \hat{p}_i)$, we take the $m$ most mixed normal modes of region $A$, which we will denote by $(\hat{Q}_i, \hat{P}_i)$, where $i=1,\dots, m$. Then, we can apply an $m$-mode swap operator,
\begin{equation}\label{nmodeswap}
    \hat{U} \coloneqq \prod_{i=1}^m \hat{U}_{\text{SWAP}}^{(i)} =  \exp\bigg[\ii\dfrac{\pi}{2}\sum_{i=1}^{m}(\hat{q}_i \hat{P}_i - \hat{Q}_i \hat{p}_i)\bigg],
\end{equation}
which acts by mapping
\begin{equation}
\begin{array}{rcl}
    \hat{q}_i&\mapsto& \hat{Q}_i\\
    \hat{Q}_i&\mapsto& -\hat{q}_i\\
    \hat{p}_i&\mapsto& \hat{P}_i\\
    \hat{P}_i&\mapsto& -\hat{p}_i
\end{array}
\end{equation}
and can be seen as generated by an interaction Hamiltonian of the form
\begin{equation}
    \hat{H}_\text{I} = \dfrac{\pi}{2}\delta(t)\sum_{i=1}^{m}\left(\hat{Q}_i\hat{p}_i - \hat{q}_i\hat{P}_i\right).
\end{equation}
Since this operation swaps the phase space of each detector degree of freedom with that of one of the normal modes of the region, all of the correlations between the chosen normal modes of $A$ and its complement $\bar{A}$ will be transmitted to the detectors. 

Recall that the quadratures of the normal modes of $A$ can be re-written as linear combinations of the quadratures of the local oscillators in $A$. The unitary in Eq.~\eqref{nmodeswap} can therefore be reinterpreted as a coupling between each detector and a privileged linear combination of the local degrees of freedom of the harmonic lattice, in the form 
\begin{equation}\label{discrete interaction}
    \hat{H}_\text{I} =\lambda\chi(t)\left(\hat{q}_\text{d}\sum_i f_i\hat{p}_i - \hat{p}_\text{d}\sum_i g_i \hat{q}_i\right),
\end{equation}
which is the discrete version of Eq.~\eqref{fieldcoupling}. In the continuum limit, the linear combination of the different spatial modes of $A$ will translate into a series of smearing functions that define the continuum limit of each normal mode (see Sec.~\ref{sectioncontinuum}).

\section{Entanglement area laws from particle detectors}\label{AreaLawSection}

In the previous section we found bounds on how much entanglement a set of $m$ detectors can acquire with the complement of the region that they couple to. The question remains of how many detectors are necessary in order to capture enough entanglement to faithfully represent the entanglement structure of the field state. In particular, we are going to analyze whether it is possible to increase the number of detectors so that they capture the entanglement area laws that the lattice theory should display, and determine the smallest number of detectors that one needs for this.

As mentioned in Sec.~\ref{DiscreteFieldSection}, subregions of the lattice theory display area law growth of entanglement with increasing size when all of the region's degrees of freedom are taken into account at each step. However, from our discussion in Sec.~\ref{OptimalCouplingSection} we know that (i) the entanglement between a region and its complement can be fully characterized by the mixedness of its normal modes, and (ii) the most mixed normal modes of a region have strong support only around the region's boundary. This suggests that it may be possible to extract a similar pattern of entanglement growth by considering only a subset of the degrees of freedom for a given region.

\begin{figure*}
    \includegraphics[width=\textwidth]{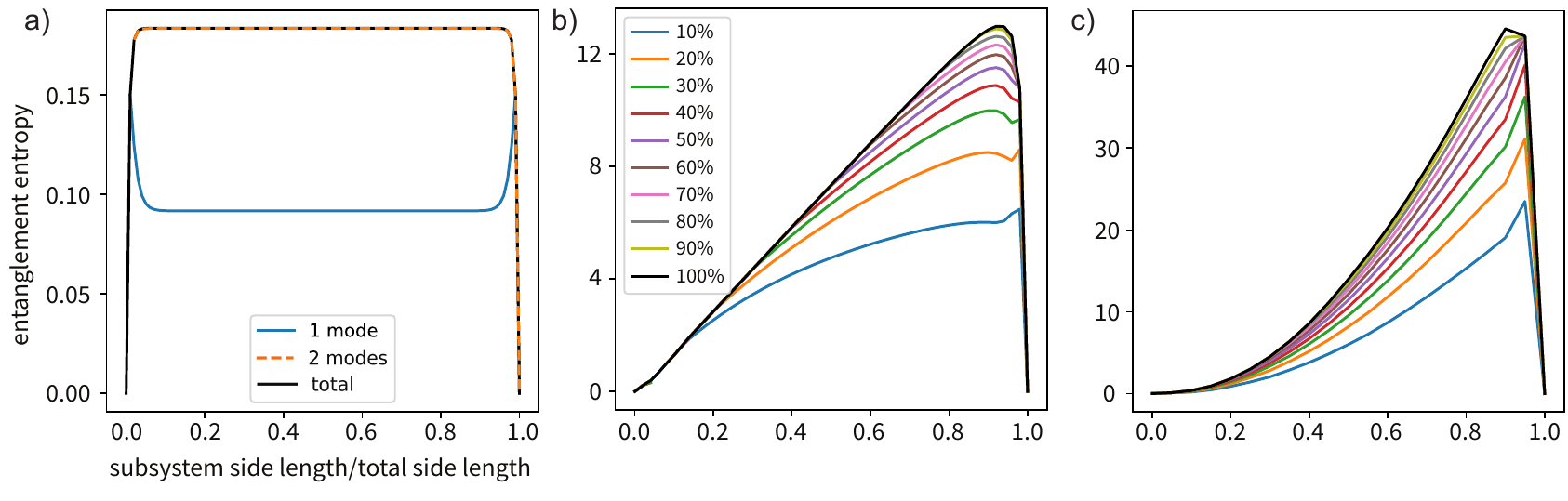}
    \caption{Entanglement entropy between a subregion and its complement as a function of the ratio between the subregion's side length and the total side length of the lattice. Subfigures a), b), and c) show this in 1, 2, and 3 spatial dimensions, respectively. If we take the number of boundary sites to be $N_\text{B}$, then each colored line corresponds to the entanglement captured by the given fraction of $N_\text{B}$ modes. The total entanglement entropy is saturated, and we observe area-law behavior in each case, when we count the entanglement coming from $N_\text{B}$ modes.}
    \label{fig:modeCounting}
\end{figure*}

This is indeed the case: if the boundary of a given region $A$ consists of $N$ sites, then one only needs the first $N$ normal modes of subregion $A$ to display area law behavior, thus encompassing nearly all of the entanglement between $A$ and its complement. Once again, we can verify this explicitly in $1$, $2$, and $3$ dimensions as shown in Fig.~\ref{fig:modeCounting}, which uses the same parameters as Fig.~\ref{fig:occupationPlots}. In $1$ spatial dimension, for any subregion size, the $2$ most mixed normal modes already encode all of the region's entanglement with the rest of the system. In higher dimensions, the number of modes has to grow at each step, but it does so more slowly than the subsystem's size: whereas the total number of degrees of freedom in a given region scales with the region's volume, the number of degrees of freedom supported at the boundary only scales with the area. Therefore, the fraction of modes needed decreases with $1/L$ as the region's side length grows. 

This fact has a direct implication on our use of detectors. Namely, it shows that, in order to capture area law behavior for the entanglement between a given region and its complement, the number of detector degrees of freedom that is needed scales with the region's boundary area. Furthermore, the optimal choice for the detector couplings would be one that can effectively be simplified to a swap of the detectors' state and the state of the most mixed normal modes of the given region. 

Notice that it is important to distinguish the entropy of a single mode $M$ within $A$ and the entropy of the most mixed normal mode in subregion $A$. Indeed, the entropy of a single mode in subregion $A$ could be larger than the entropy of the most mixed normal mode in $A$. However, this does not mean that $M$ has more entanglement with the complement of $A$ than the most mixed normal mode. Instead what happens is that $M$ is entangled both with other modes in $A$, as well as possibly modes in the complement of $A$. A normal mode of $A$ does not have any entanglement with other normal modes in $A$, and therefore its entropy quantifies entanglement with modes that live entirely in $\bar A$.

This is important when working with particle detectors, since the interaction of the detector with the lattice itself can create extra entanglement between modes in the interaction region. For instance, if one does not work with normal modes, it is always possible to choose one single mode supported on a given region such that the von Neumann entropy of this mode equals the full entanglement entropy between the region and its complement. One simple way to do this is to pick one of the modes obtained from a two-mode squeezing between the two most mixed normal modes of subregion $A$. One of the modes thus generated will always have a higher von Neumann entropy than the most mixed normal mode of $A$, and by making the squeezing parameter high enough, such von Neumann entropy can be made equal to the total region's von Neumann entropy. By then swapping this mode with just one single-mode detector, the detector would acquire a von Neumann entropy that would also scale according to an area law. It is clear, however, that the resulting mixedness of the detector in the final state would not correspond to entanglement with the complement of the coupling region. On the contrary, the entanglement in this case would be mostly shared with a mode fully supported in the region the detector couples to, and would be almost entirely generated by the direct interaction of the detector and the lattice. 

Similarly, a naive swap between $N$ detector degrees of freedom and the $N$ local oscillators supported on the boundary of the coupling region would not serve our purposes. This too would lead to a von Neumann entropy that scales with the area of the subregion of interest. However, each boundary site is equally entangled with its nearest neighbours to the inside and the outside of the coupling region. Therefore, the resulting entanglement transmitted to the detectors in this case would not be representative of the entanglement between the coupling region and its complement either.

\section{The continuum limit}\label{sectioncontinuum}

The results of the previous sections have been obtained using harmonic lattices as a discrete approximation of a quantum field theory. In this section, we will prescribe how to generalize the results when we take the continuum limit in which the lattice theory becomes a quantum field theory in a box of finite fixed length. In order to do that, we will analyze the opposite procedure, i.e., the discretization that takes us from the continuum to the lattice. This will reveal how to implement the limit carefully, and it will help to understand the subtleties behind the discrete approximation made in Section~\ref{DiscreteFieldSection}.

Consider a real scalar field theory constrained to a volume $V=[0,L]^n \in \mathbb{R}^{n}$ with periodic boundary conditions, and with Hamiltonian
\begin{equation}\label{fieldhamiltonianbox}
    \hat{H}_{\phi} = \dfrac{1}{2}\int_{V} \dd^n \bm{x}\left(\hat{\pi}^2 + (\nabla \hat{\phi})^2 + m^2\hat{\phi}^2\right).
\end{equation}
The periodic boundary conditions allow us to expand the field in a Fourier series,
\begin{equation}\label{field in a box}
\hat\phi(t,\bm{x})=\frac{1}{L^n} \sum_{\bm{m} \in \mathbb{Z}^{n}} \hat{\varphi}(t,\bm{k}_{\bm{m}})\, e^{-\ii\bm{k}_{\bm{m}}\cdot\bm{x}},
\end{equation}
where $\bm{k}_{\bm{m}}=\frac{2\pi}{L}\bm{m}$. We can now impose a UV cutoff by setting that $\hat\varphi(t,\bm{k}_{\bm{m}})=0$ for any $\bm{m}$ such that its largest component in absolute value, $||\bm{k}_{\bm{m}}||_\infty$, is bigger than $K$, with \mbox{$K=\pi (2N+1)/L$}, for a positive integer $N\geq 1$. This allows us to write
\begin{equation}\label{field in a box UV}
\hat\phi(t,\bm{x})=\frac{1}{L^n}\sum_{\bm{m}} \hat\varphi(t,\bm{k_m}) \, e^{-\ii \bm{k_m} \cdot \bm{x}} ,
\end{equation}
where the sum runs over $\bm{m} \in \mathbb{Z}^{n} \cap [-N,N]^n$. In particular, for $\bm{l} \in \mathbb{Z}^{n} \cap [0,2N]^n$, we have that
\begin{equation}
\hat{\phi}\bigg(t,\frac{L}{2N+1}\bm{l} \bigg)=\frac{1}{L^n}\sum_{\bm{m}} \hat\varphi(t,\bm{k_m})\,e^{-\frac{2\pi\ii}{2N+1}\bm{m}\cdot\bm{l}},
\end{equation}
which means that 
\begin{equation}
\bigg\{ \hat{\phi}\bigg(t,\frac{L}{2N+1}\bm{l} \bigg) \,:\, \bm{l} \in \mathbb{Z}^{n} \cap [0,2N]^n \bigg\}
\end{equation}
and
\begin{equation}
\bigg\{\frac{1}{L^n}\hat\varphi(t,\bm{k_m}) \,:\,\bm{m} \in \mathbb{Z}^{n} \cap [-N,N]^n \bigg\} 
\end{equation}
are related by a discrete Fourier transform. Thus,
\begin{equation}
\hat\varphi(t,\bm{k_m})=a^n \sum_{\bm{l}} \hat\phi(t,a\bm{l})\,e^{\frac{2\pi\ii}{2N+1}\bm{m}\cdot\bm{l}},
\end{equation}
where $a=\pi/K=L/(2N+1)$. By Eq.~\eqref{field in a box}, the field can be expressed as
\begin{equation}
\hat\phi(t,\bm{x})=\frac{1}{(2N+1)^n}\sum_{\bm{m}}\sum_{\bm{l}} \hat\phi(t,a\bm{l})\,e^{-2\pi\ii\bm{m}\cdot(\bm{x}-a\bm{l})/L},
\end{equation}
where the sums are running over \mbox{$\bm{m}\in[-N,N]^n$} and \mbox{$\bm{l}\in[0,2N]^n$}. The sum in $\bm{m}$ can be evaluated to yield
\begin{equation}\label{discrete field}
\hat\phi(t,\bm{x})=\sum_{\bm{l}} \hat\phi(t,a\bm{l})\, \mathcal{S}^{(n)}_{\bm{l}}\Big(N;\frac{\bm{x}}{a}\Big),
\end{equation}
where we have denoted
\begin{equation}
\mathcal{S}^{(n)}_{\bm{l}}\Big(N;\frac{\bm{x}}{a}\Big)\coloneqq\prod_{r=1}^{n}\mathcal{S}_{l_r}\Big(N;\frac{x_r}{a}\Big),
\end{equation}
with $l_r$ and $x_r$ denoting the $r$-th component of $\bm{l}$ and $\bm{x}$, respectively, and
\begin{equation}
\mathcal{S}_j\Big(N;\frac{y}{a}\Big)\coloneqq \frac{\sin\big(\pi\big[\frac{y}{a}-j\big]\big)}{(2N+1)\sin\big(\frac{\pi}{2N+1}\big[\frac{y}{a}-j\big]\big)}.
\end{equation}
The same procedure can be applied to the momentum $\hat\pi(t,\bm{x})$ and the gradient $\nabla\hat\phi(t,\bm{x})$ yielding
\begin{equation}\label{discrete field momentum}
\hat\pi(t,\bm{x})=\sum_{\bm{l}} \hat\pi(t,a\bm{l})\, \mathcal{S}^{(n)}_{\bm{l}}\Big(N;\frac{\bm{x}}{a}\Big),
\end{equation}
and
\begin{equation}\label{discrete field gradient}
\nabla\hat\phi(t,\bm{x})=\sum_{\bm{l}} \nabla\hat\phi(t,a\bm{l})\, \mathcal{S}^{(n)}_{\bm{l}}\Big(N;\frac{\bm{x}}{a}\Big).
\end{equation}
The conjunction of the IR cutoff imposed by the box constraint and the UV cutoff allows us to reduce the description of the field in the continuum to a finite number of degrees of freedom, namely, those of the field at the sampling positions $\{a\bm{l},\,\bm{l}\in[0,2N]^n\}$. 
It is worth pointing out that this approach is related to performing a wavelet discretization of the QFT~\cite{Wavelet1}. Indeed similar techniques have in fact been used in the past in related contexts (see for instance~\cite{Wavelet2, Wavelet3}).  

Once we have reduced the system to a finite number of degrees of freedom, we can use Eqs.~\eqref{discrete field},~\eqref{discrete field momentum},~\eqref{discrete field gradient} to obtain a discrete version of the Hamiltonian~\eqref{fieldhamiltonianbox}:
\begin{align}
\label{field discrete 1}
    \hat{H}_\phi=&\frac{1}{2}\sum_{\bm{m},\bm{l}} (\hat\pi(t,a\bm{m})\,\hat\pi(t,a\bm{l})+\nabla\hat\phi(t,a\bm{m})\,\nabla\hat\phi(t,a\bm{l}) \nonumber \\
    &\phantom{===\,}+m^2\,\hat\phi(t,a\bm{m})\,\hat\phi(t,a\bm{l}))\,\mathcal{I}_{\bm{m}\bm{l}}\,,
\end{align}
where
\begin{align}
    \mathcal{I}_{\bm{m}\bm{l}}&=\int_V \dd^{n}\bm{x}\;\mathcal{S}^{(n)}_{\bm{m}}\Big(N;\frac{\bm{x}}{a}\Big) \,\mathcal{S}^{(n)}_{\bm{l}}\Big(N;\frac{\bm{x}}{a}\Big) \\
    &=\prod_{r=1}^{n} \int_{0}^{L}\dd x_{r} \;\mathcal{S}_{m_r}\Big(N;\frac{x_r}{a}\Big) \,\mathcal{S}_{l_r}\Big(N;\frac{x_r}{a}\Big) \\
    &=a^n \prod_{r=1}^{n}\delta_{m_r l_r}=a^n \,\delta^{(n)}_{\bm{ml}}\,. \label{I_lm}
\end{align}
Since these integrals cancel out when $\bm{m}\neq\bm{l}$, the Hamiltonian reduces to
\begin{align}
\label{field discrete 1 final}
    \hat{H}_\phi=&\frac{a^n}{2}\sum_{\bm{m}} (\hat\pi(t,a\bm{m})^2+\nabla\hat\phi(t,a\bm{m})^2+m^2\,\hat\phi(t,a\bm{m})^2).
\end{align}
Notice that in Eq.~\eqref{field discrete 1 final} we still have derivatives of the field, so even though we reduced the description of the field to a finite number of degrees of freedom, we still rely on the field being defined in a continuum. To complete the discretization and reduce the QFT to a lattice of harmonic oscillators, we need to discretize the derivative. Let us consider the partial derivative in the direction of the $s$-th coordinate axis
\begin{align}
    \partial_s \hat\phi(t,\bm{x})&=\sum_{\bm{m}}\partial_s\hat\phi(t,a\bm{m})\,\mathcal{S}^{(n)}_{\bm{m}}\Big(N;\frac{\bm{x}}{a}\Big) \\
    &=\sum_{\bm{m}}\hat\phi(t,a\bm{m})\, \partial_s \mathcal{S}^{(n)}_{\bm{m}}\Big(N;\frac{\bm{x}}{a}\Big).
\end{align}
Thus, we can write
\begin{equation}
    \sum_{\bm{m}}\partial_s \hat\phi(t,a\bm{m})\, \mathcal{I}_{\bm{m}\bm{l}}=\sum_{\bm{m}} \hat\phi(t,a\bm{m}) \,\mathcal{J}_{\bm{ml}}^{s},
\end{equation}
where
\begin{equation}
    \mathcal{J}_{\bm{ml}}^s=\int_{V}\dd^n\bm{x}\, \partial_s \mathcal{S}^{(n)}_{\bm{m}}\Big(N;\frac{\bm{x}}{a}\Big)\,\mathcal{S}^{(n)}_{\bm{l}}\Big(N;\frac{\bm{x}}{a}\Big).
\end{equation}
It can be shown that
\begin{align}
    \mathcal{J}^s_{\bm{ml}}=&(1-\delta_{m_s l_s})\Bigg[\frac{(-1)^{l_s-m_s}}{l_s-m_s}a^{n-1}\prod_{r\neq s} \delta_{m_r l_r} +\mathcal{O}(N^{-2})\Bigg]. 
\end{align}
Thus, when $N\gg 1$,
\begin{align}
\label{approx derivative high N}
    \partial_s \hat\phi(t,a\bm{l})=\frac{1}{a}\sum_{\bm{m}}&\frac{(-1)^{l_s-m_s}}{l_s-m_s}(1-\delta_{l_s m_s}) \\
    &\times\prod_{r\neq s} \delta_{l_r m_r}\, \hat\phi(t,a\bm{m})+\mathcal{O}(N^{-2}). \nonumber
\end{align}
This allows us to approximate the gradient $\nabla \hat\phi(t,a\bm{l})$ in terms of the field amplitude at the sampling points, finally completing the discretization, i.e., the reduction of the field system to a lattice of harmonic oscillators. Since the expression for the derivative at one point includes the field amplitude at different sites, the term $\nabla \hat\phi(t,a\bm{l})^2$ in Eq.~\eqref{field discrete 1 final} leads to couplings between the amplitudes at different sites, and therefore we see explicitly that the price we pay for imposing a UV cutoff is to have a non-local Hamiltonian, i.e., we need to give up locality. However, if we additionally  perform an approximation of the derivative to nearest neighbours (see, for instance, \cite{Dan2022Part2}), we get that the field Hamiltonian reduces to 
\begin{align}
    \nonumber\hat H_\phi \approx\;\;&\frac{a^n}{2} \sum_{\bm{l}} \Bigg( \hat\pi(t,a\bm{l})^2+ m^2\hat\phi(t,a\bm{l})^2 \\
    &+\sum_{r=1}^{n} \Bigg[ \frac{\hat\phi(t,a(\bm{l}+\bm{1}_r))-\hat\phi(t,a\bm{l})}{a} \Bigg]^2\,\Bigg), 
\end{align}
where $\bm{1}_r$ is the unit vector with all components equal to zero except for the $r$-th component. Finally, defining
\begin{equation}\label{discrete parameters}
\omega^2=m^2+\frac{2n}{a^2}, \quad \alpha=\frac{1}{\omega a^2},
\end{equation}
and the dimensionless quadratures
\begin{equation}
\label{discrete quadratures}
    \hat q_{\bm{m}}(t)=\sqrt{\omega a^n} \, \hat\phi(t,a\bm{m}), \quad \hat p_{\bm{m}}(t)=\sqrt{\frac{a^n}{\omega}}\,\hat\pi(t,a\bm{m}),
\end{equation}
we recover Eq.~\eqref{discretefieldhamiltonian}. Notice that by setting $a=\varepsilon$ in Eqs.~\eqref{discrete parameters} and~\eqref{discrete quadratures} we recover Eqs.~\eqref{latticefrequency}--\eqref{latticecoupling} from Section~\ref{DiscreteFieldSection}. 

\begin{figure}[H]
    \includegraphics[scale=.92]{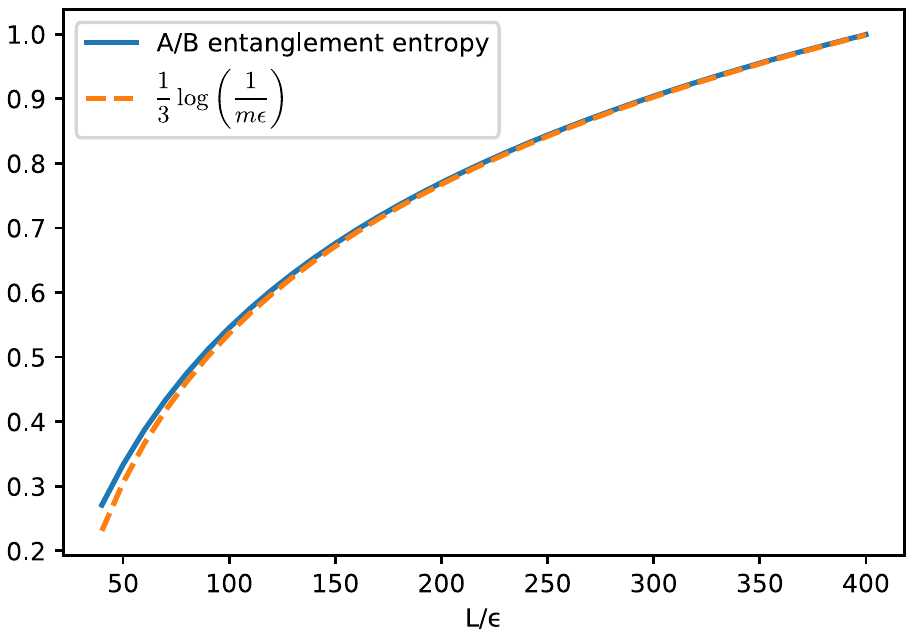}
    \caption{Entanglement entropy between one third of a 1D chain and its complement as a function of the total number of sites in the lattice, i.e. the inverse UV cutoff, as here we keep the length of the chain fixed. As the density of lattice sites increases, we see that the dependence of the entanglement entropy on the effective short distance cutoff quickly converges to the behavior that is expected from other methods in field theory~\cite{EntropyHolographyReview}.}
    \label{fig:modeCounting2}
\end{figure}

The analysis of the discretization procedure hints at how to go from a discrete lattice to the continuum. If we consider a $n$-dimensional hypercubic lattice of $(2N+1)^n$ sites  with dimensionless quadratures $(q_{\bm{l}},p_{\bm{l}})$ and Hamiltonian given by Eq.~\eqref{discretefieldhamiltonian}, given a fixed length $L$, we can define 
\begin{equation}
    \hat\phi_{N}(t,\bm{x})=\sum_{\bm{l}}\frac{1}{\sqrt{\omega a^n}} \hat q_{\bm{l}} \,\mathcal{S}^{(n)}_{\bm{l}}\Big( N;\frac{\bm{x}}{a} \Big),
\end{equation}
\begin{equation}
    \hat\pi_{N}(t,\bm{x})=\sum_{\bm{l}}\sqrt{\frac{\omega}{a^n}} \hat p_{\bm{l}} \,\mathcal{S}^{(n)}_{\bm{l}}\Big( N;\frac{\bm{x}}{a} \Big),
\end{equation}
for $\bm{x}\in[0,L]^{n}$, and
\begin{equation}
    \hat{H}_{\phi_N} = \dfrac{1}{2}\int_{V} \dd^n \bm{x}\left(\hat{\pi}_N^2 + (\nabla \hat{\phi}_N)^2 + m^2\hat{\phi}_N^2\right).
\end{equation}
where $a=L/(2N+1)$, and $\omega$ is defined as in Eq.~\eqref{discrete parameters}. In the limit $N\to \infty$,
\begin{equation}
    \hat\phi_N(t,\bm{x}) \to \hat\phi(t,\bm{x}), \quad \hat\pi_N(t,\bm{x}) \to \hat\pi(t,\bm{x}),
\end{equation}
and 
\begin{equation}
    \hat{H}_{\phi_{N}}(t,\bm{x}) \to \hat{H}_{\phi}(t,\bm{x}).
\end{equation}
If one were to claim that a particular theory with first-neighbours interactions on the lattice is a good approximation to a QFT in the continuum one needs to make sure that two different approximations hold: 1) that  $N$ is large enough for the discrete field theory to approximate the continuum theory and 2) that the discrete representation of the derivative can be approximated well enough by a first neighbours finite difference. For our purposes we can quickly check that the lattice results for entanglement entropy in the lattice quickly converge to the expected behavior in the continuum as $N$ is increased (see for example Fig.~\ref{fig:modeCounting2}).

\begin{figure*}
    \includegraphics[width=\textwidth]{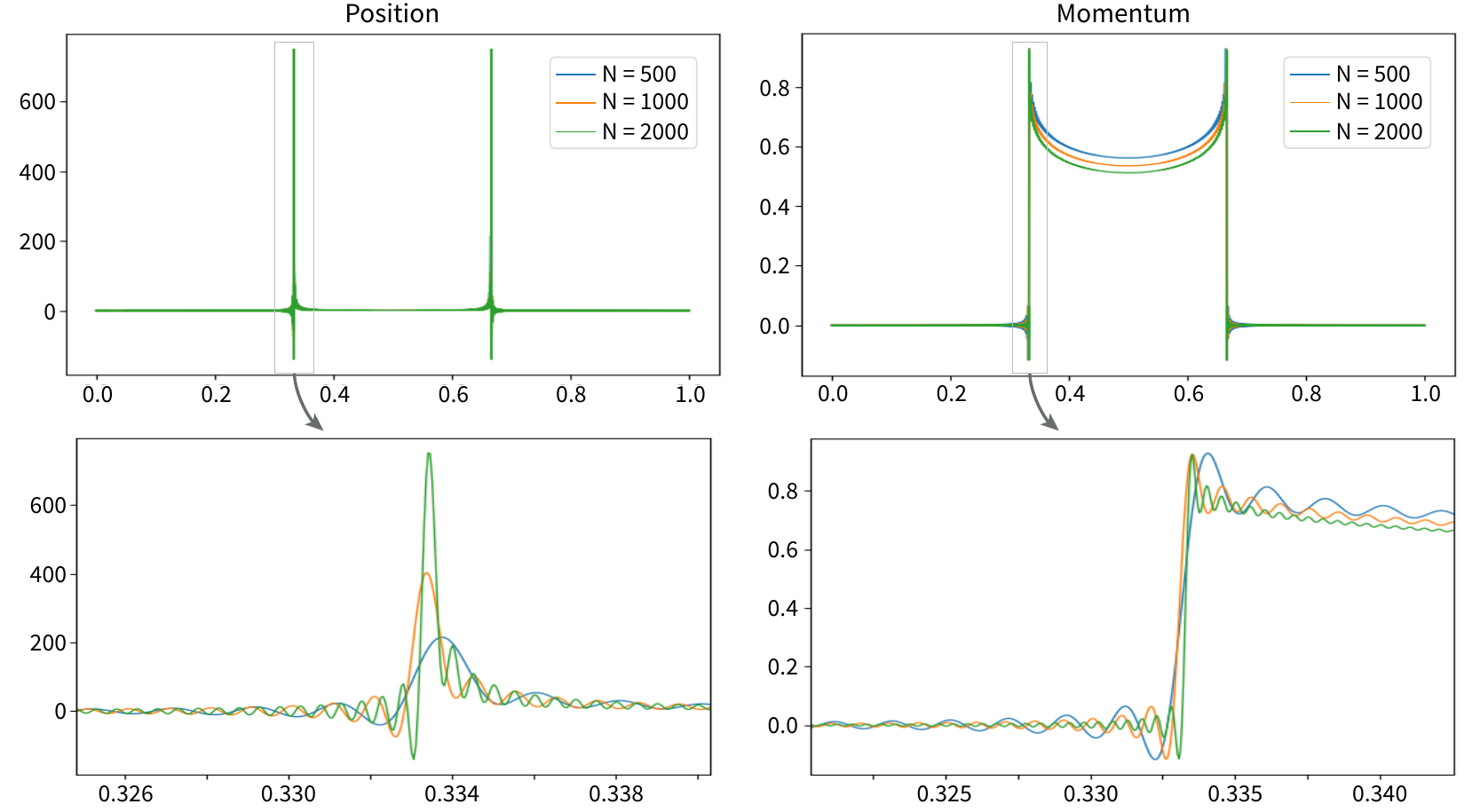}
    \caption{Dimensionless approximated position and momentum profiles, $\lambda_p g_{N}(x)$ and $\lambda_q f_{N}(x)$, of the most mixed mode of a region comprising the central third of a box, for a 1+1-dimensional massive quantum scalar field theory of mass $m=1/L$ in the vacuum state. The profiles were calculated using Eqs.~\eqref{approximate momentum profile} and~\eqref{approximate position profile}, using the spatial profiles obtained for the corresponding 1-dimensional lattice field theories with UV parameters $N=500$, $1000$ and $2000$, and setting the parameters $\omega$ and $\alpha$ via Eq.~\eqref{discrete parameters} with $n=1$. The second row shows an enlarged view of the profiles close to the left boundary.}
    \label{fig:1D}
\end{figure*}

Regarding the interaction between the field and the detector, in the continuum it is described by the Hamiltonian given in Eq.~\eqref{fieldcoupling}. When we consider the discretized expressions for the field~\eqref{discrete field} and the momentum~\eqref{discrete field momentum}, Eq.~\eqref{fieldcoupling} reduces to
\begin{align}
\label{discretefieldcoupling}
    \hat{H}_{\text{I}} =& \lambda\chi(t)\bigg(\lambda_q\,\hat{q}_d\,\sum_{\bm{l}}\hat\pi(t,a\bm{l})\int_V\dd^n \bm{x}\, f(\bm{x})\,\mathcal{S}^{(n)}_{\bm{l}}\Big(N;\frac{\bm{x}}{a}\Big) \nonumber \\
    &- \lambda_p\,\hat{p}_d\,\sum_{\bm{l}}\hat\phi(t,a\bm{l})\int_V\dd^n \bm{x}\, g(\bm{x})\,\mathcal{S}^{(n)}_{\bm{l}}\Big(N;\frac{\bm{x}}{a}\Big)\bigg).
\end{align}
Thus, to recover Eq.~\eqref{discrete interaction}, we simply define
\begin{equation}
\label{f discrete}
    f_{\bm{l}}=\lambda_q \sqrt{\frac{\omega}{a^n}}\int_V\dd^n \bm{x}\, f(\bm{x})\,\mathcal{S}^{(n)}_{\bm{l}}\Big(N;\frac{\bm{x}}{a}\Big)
\end{equation}
and
\begin{equation}
\label{g discrete}
    g_{\bm{l}}=\frac{\lambda_p}{\sqrt{\omega a^n}}\int_V\dd^n \bm{x}\, g(\bm{x})\,\mathcal{S}^{(n)}_{\bm{l}}\Big(N;\frac{\bm{x}}{a}\Big)
\end{equation}
Conversely, to recover the spatial smearing functions $\{f(\bm{x}),g(\bm{x})\}$ in the continuum from the  spatial profiles on the lattice $\{f_{\bm{l}},g_{\bm{l}}\}$, we can define
\begin{equation}
    \label{approximate momentum profile}
    f_{N}(\bm{x})=\frac{1}{\lambda_q\sqrt{\omega a^n}}\sum_{\bm{l}}f_{\bm{l}}\,\mathcal{S}^{(n)}_{\bm{l}}\Big(N;\frac{\bm{x}}{a}\Big),
\end{equation}
and
\begin{equation}
\label{approximate position profile}
    g_{N}(\bm{x})=\frac{1}{\lambda_p}\sqrt{\frac{\omega}{ a^n}}\sum_{\bm{l}}g_{\bm{l}}\,\mathcal{S}^{(n)}_{\bm{l}}\Big(N;\frac{\bm{x}}{a}\Big).
\end{equation}
By Eq.~\eqref{I_lm}, one can see from Eqs.~\eqref{f discrete} and~\eqref{g discrete} that
\begin{equation}
    f_{N}(\bm{x}) \to f(\bm{x}) \quad \textrm{and} \quad g_{N}(\bm{x}) \to g(\bm{x})
\end{equation}
when $N \to \infty$, and thus the modes $\hat Q$ and $\hat P$ of Eq.~\eqref{defnormalmodes} converge to those of Eqs.~\eqref{defnormalmodescontinuumQ} and~\eqref{defnormalmodescontinuumP}. Note that all the convergences so far stated are achieved pointwise in general, rather than uniformly.

\begin{figure*}[]
    \includegraphics[width=\textwidth]{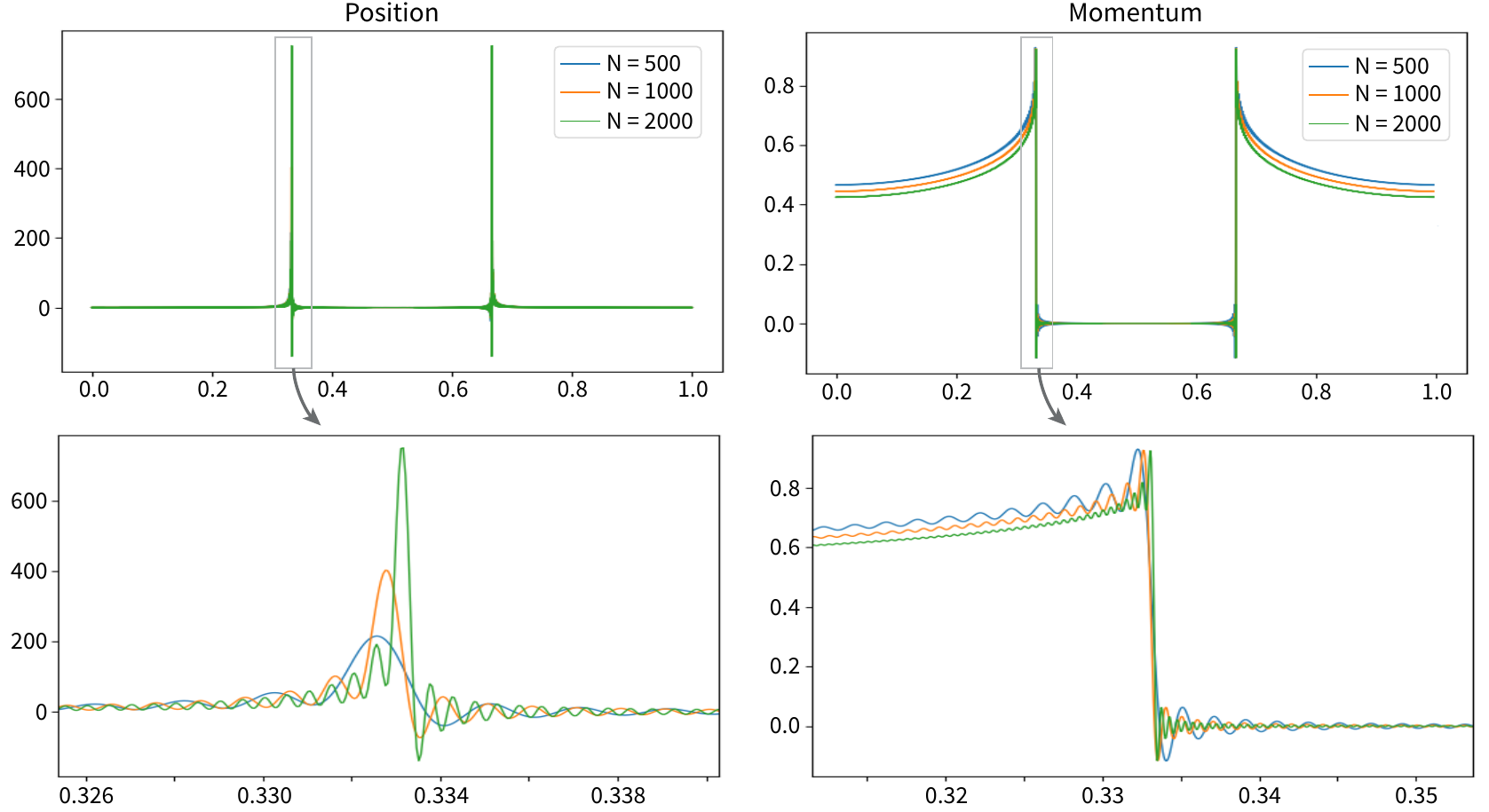}
    \caption{Dimensionless approximated position and momentum profiles, $\lambda_p g_{N}(x)$ and $\lambda_q f_{N}(x)$, of the partner mode of the most mixed mode of a region comprising the central third of a box, for a 1+1-dimensional massive quantum scalar field theory of mass $m=1/L$ in the vacuum state. The profiles were calculated using Eqs.~\eqref{approximate momentum profile} and~\eqref{approximate position profile}, using the spatial profiles obtained for the corresponding 1-dimensional lattice field theories with UV parameters $N=500$, $1000$ and $2000$, and setting the parameters $\omega$ and $\alpha$ via Eq.~\eqref{discrete parameters} with $n=1$. Below the profiles, a zoom in of the detail of the profiles close to the left boundary.}
    \label{fig:1Dcompl}
\end{figure*}

We can use these results to analyze what happens to the position and momentum profiles of the normal modes of a region and its complement when one approaches the continuum limit. In particular, here we will examine the most mixed normal mode. 

In Fig.~\ref{fig:1D} we see that, in 1D, as we increase the UV cutoff---and therefore the number of sites of the underlying harmonic chain---the momentum profile $f(x)$ of the most mixed mode of a region formed by a third of the total number of sites seems to converge to a bounded function which is peaked at the boundary and increasingly suppressed within the interior of the region. Meanwhile, the position profile $g(x)$ gets increasingly sharper at the boundary and is almost zero in the interior, hinting at a convergence towards the even sum of two delta functions supported in the two boundary points. Fig.~\ref{fig:1Dcompl} shows the same pattern for its partner mode in the complementary region of the chain. We also see that as we approach the continuum limit, the momentum profile localization for the most mixed mode in region $A$ reaches a local minimum in the centre of the region. For its partner mode, the momentum profile reaches a minimum in the centre of the complement region. We cannot verify numerically whether there is a lower bound for this minimum, but a more detailed study of the behavior of this profile in the continuum limit should be worth exploring in the future. 

\begin{figure*}[]
    \includegraphics[width=\textwidth]{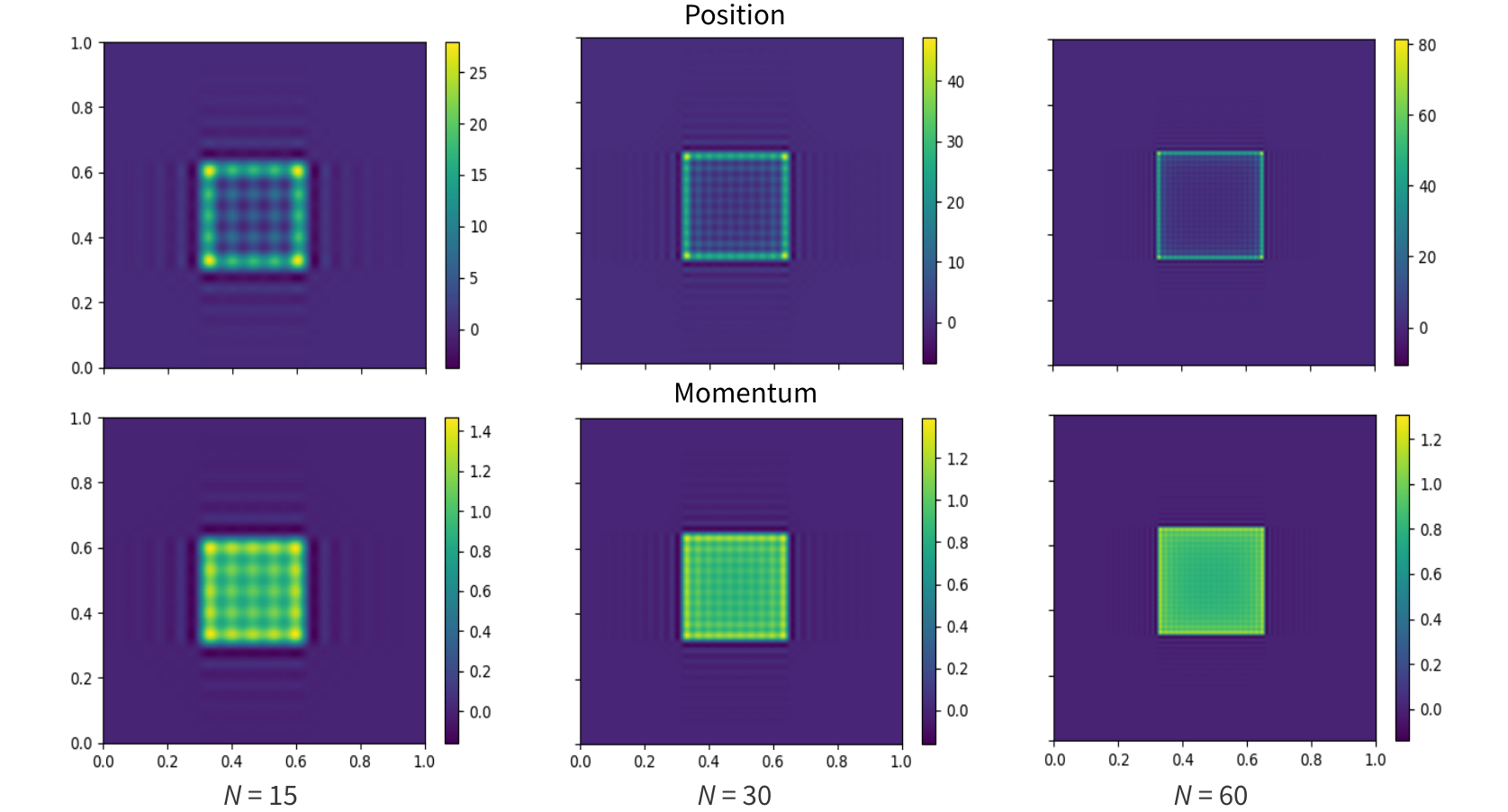}
    \caption{Dimensionless approximated position and momentum profiles, $\lambda_p g_{N}(x)$ and $\lambda_q f_{N}(x)$, of the most mixed mode of a region comprising the central square of side length $L/3$ of a square of side length $L$, for a 2+1-dimensional massive quantum scalar field theory of mass $m=1/L$ in the vacuum state. The profiles were calculated using Eqs.~\eqref{approximate momentum profile} and~\eqref{approximate position profile}, using the spatial profiles obtained for the corresponding 2-dimensional lattice field theories with UV parameters $N=15$, $30$ and $60$, and setting the parameters $\omega$ and $\alpha$ via Eq.~\eqref{discrete parameters} with $n=2$.}
    \label{fig:2D}
\end{figure*}

\begin{figure*}[]
    \includegraphics[width=\textwidth]{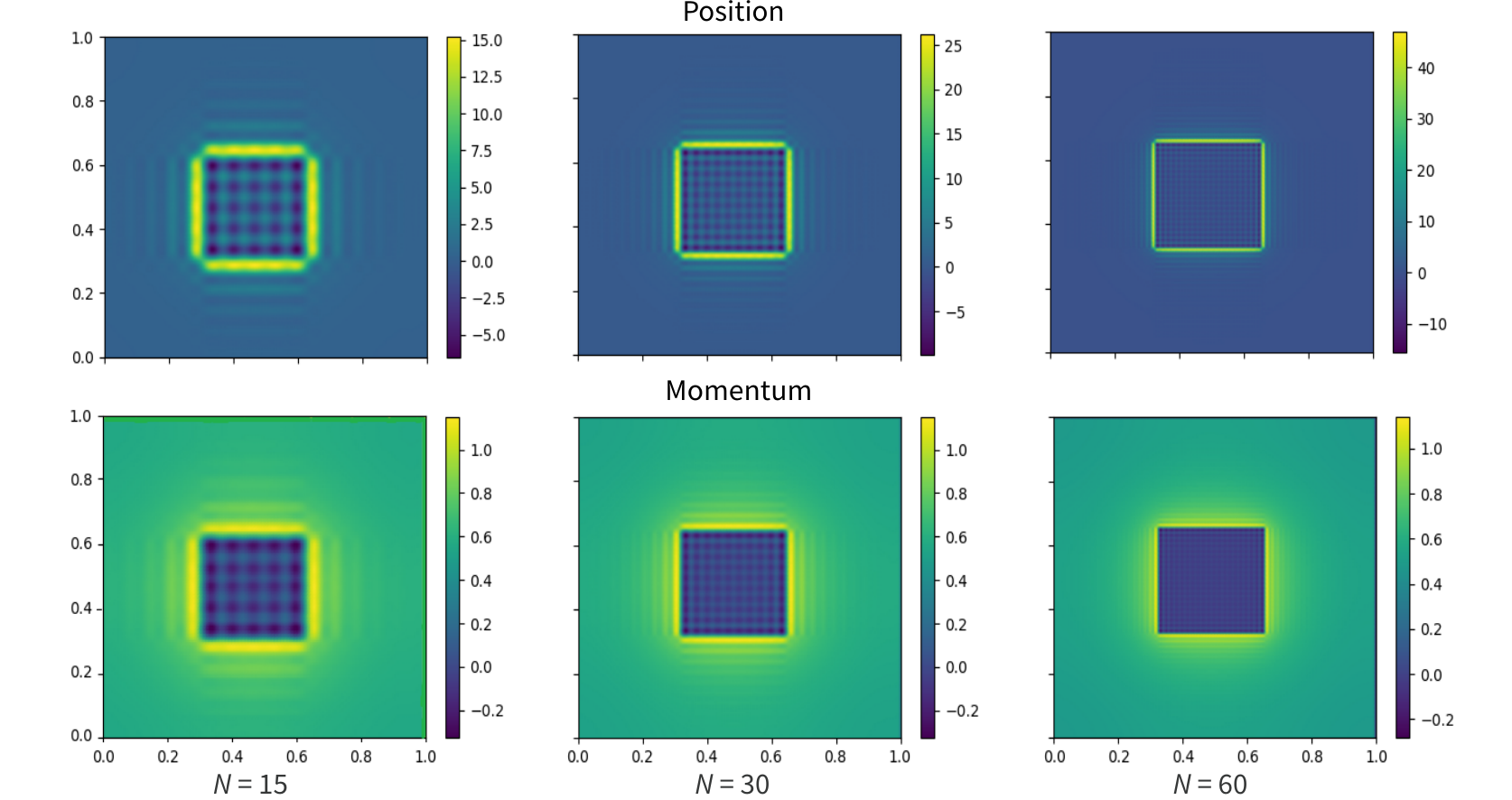}
    \caption{Dimensionless approximated position and momentum profiles of the partner mode of the mode represented in Fig.~\ref{fig:2D}.}
    \label{fig:2Dcompl}
\end{figure*}

The behavior displayed in 1D seems to extend to 2D. Following the parameters of the 1D case, we considered a square region with a side length of a third of that of the lattice. In Fig.~\ref{fig:2D}, we observe that all the approximated momentum profiles are bounded, peak at the boundary and decrease towards the interior. Meanwhile, the position profiles peak at the boundary with increasingly high maxima, and are all approximately zero in the interior, hinting again at a convergence to a delta at the boundary. A similar behavior is again displayed by the partner mode in the complementary region (see Fig.~\ref{fig:2Dcompl}).

Notice that the convergence of sequences of continuous functions to delta distributions is yet another evidence of the fact pointed out before that these convergences are pointwise and not uniform in general. Another detail worth mentioning is that the approximated profiles in position and momentum show regions in which they are negative. These only show up outside of the support that the profiles are expected to have in the continuum limit. This an instance of the well-known Gibbs phenomenon (see, e.g.,~\cite{Jerri1998}), and it is a natural consequence of approximating with Fourier sums functions that have jump discontinuities, as the exact profiles are expected to have in the boundary. A simple way to overcome this issue when using the approximate profiles in practice is to set them to zero by fiat out of their known support.

\section{Discussion}\label{ConclusionSection}

Measuring entanglement between arbitrary spacetime regions in a field theory is still an open problem. Inspired by the protocols of entanglement harvesting~\cite{Valentini1991,Reznik2003,Reznik2,Pozas-Kerstjens:2015}, we proposed here that particle detectors (localized quantum systems that couple to quantum fields) can provide a natural way to accurately quantify this entanglement without running into ill-defined quantities due to UV divergences or complexity problems in the evaluation of entanglement monotones.

In order for this proposal to be viable, one needs to show that particle detectors can indeed faithfully capture the entanglement that the region they couple to has with other regions of spacetime. Just applying entanglement harvesting protocols is not enough for this: the entanglement harvested by a pair of detectors may be a witness for the entanglement structure, but it is by no means a quantifier for the actual amount of entanglement present in the field. In general we do not know how efficiently the detectors are harvesting entanglement from the field. To know this, one needs to figure out what kinds (if any) of couplings between detectors and field allow for a faithful transfer of field entanglement to the detectors so that the entanglement contained in the detectors accurately reflects the entanglement previously present in the field. This is the question we began to address in this paper. 

In this paper, we have laid out a criterion to select, for any given subregion of a harmonic lattice, the set of modes that best encode the entanglement properties of that subregion with its complement. We then argued that, by engineering a suitable coupling between the given subregion and a set of probes, it is possible to completely transfer that entanglement to the probe degrees of freedom. This establishes the optimal form of coupling that a probe must have in order to most efficiently capture the entanglement between the coupling region and the rest of space. With this we have the ingredients needed to substitute the problem of computing entanglement measures in a many-body system with the much simpler problem of evaluating entanglement in probes coupled to the system.

As a proof of principle, we showed how this procedure allows us to reproduce one of the fundamental properties of the entanglement in the ground state of a harmonic lattice---namely, the area law growth of entanglement entropy---with only a subset of the modes in the region. The minimum number of required modes corresponds to a monotonically decreasing fraction of the total number of modes as the region size grows. In terms of detectors, this sets the minimum required number of degrees of freedom that must be accessible to probes in order for them to recover the full entanglement that was originally present in a given region. It also sets the limits on how much of the entanglement can be extracted when the probes are limited to a fixed number of degrees of freedom, which is perhaps the most realistic scenario encountered in typical setups in relativistic quantum information.

We started our analysis with an approximation to the free scalar field as a lattice of coupled harmonic oscillators with finitely many degrees of freedom in any finite region of space, which can be obtained by placing a UV cutoff in the theory. However, we also showed how to carefully extrapolate the form of the optimal coupling from the lattice theory to the continuum, which allows one to reconstruct the spatial profile of any given mode of the field as the cutoff is sent to zero. This yields the position and momentum smearing functions one would like to implement in order to couple a particle detector to a field theory fully in the continuum. From a particle-detector perspective, it is worth pointing out that intuitively, particle detectors---which generally come equipped with an inherent energy gap and a minimum finite spatial extension---provide a natural mechanism for an effective UV cutoff. We only expect detectors to be physically able to probe a finite number of modes within a given spatial region, and therefore, they can be used to define an operationally motivated cutoff to the field theory. This suggests that, from the point of view of detectors, even a continuous field theory with infinitely many degrees of freedom may behave as an effectively discretized chain of oscillators, up to the resolution allowed by the detectors~\cite{DanIreneML,Dan2022Part1,Dan2022Part2}. 

Additionally, though this analysis addressed entanglement between a region and its complement, finding normal modes and computing entanglement does not become fundamentally more difficult when the regions are of arbitrary shape, as using particle detectors of arbitrary geometry is commonplace in the literature of entanglement harvesting. Considering this, the methods developed here could open a new avenue for the exploration of the entanglement structure of a QFT (arguably, a difficult enterprise~\cite{Agullo}) in a way that is well-motivated from an operational point of view, and also naturally devoid of some of the issues that plague other approaches in field theory.

A much more challenging task is understanding how to optimally extract entanglement between non-complementary subregions of space when one cannot probe all the degrees of freedom in each region. This is the case of most interest for endeavors in entanglement harvesting. In these cases, one usually wishes to capture correlations between causally disconnected regions of spacetime by probing the field with simple physical systems that have a small number of degrees of freedom. Prescribing the optimal profile of the coupling for entanglement extraction between disjoint regions poses new challenges: if two subregions $A$ and $B$ are not complementary, the joint state of $AB$ will be mixed, and therefore the key techniques used in Sec.~\ref{OptimalCouplingSection} to select a privileged set of modes supported in the regions of interest will no longer apply in general. A first tentative conjecture could be that, if the two parties (say, Alice and Bob) wishing to extract the maximum possible entanglement only have access to one mode in each region, then the best possible choice of coupling would be to have Alice and Bob swap with the most mixed normal mode of $A$ and $B$ respectively. This may seem promising at a superficial level, since it reduces to our result in Sec.~\ref{OptimalCouplingSection} when $B$ becomes the complement of $A$; however, it is possible to find simple counter-examples to this tentative recipe, so the true answer must be more subtle.

Imagine, for instance, an overall pure Gaussian state partitioned into three regions $ABC$, such that $A$ contains two modes $A_1 A_2$, and $B$ and $C$ each have just one mode. Let us assume that $A_1$ and $A_2$ are the normal modes of subregion $A$, and that we have the partner-mode pairs $A_1C$ and $A_2B$ with the entanglement between $A_1C$ being higher than that between $A_2B$. If we take the subregion $A$, its most mixed normal mode would then be the mode $A_1$. Now, if we wish to optimally extract pairwise entanglement between the subregions $A$ and $B$, the proposal laid out in the previous paragraph would tell us to take the most mixed normal mode of $A$ (namely, the mode $A_1$) and consider its entanglement with mode $B$; but by construction these two modes are fully decoupled, and therefore the entanglement between them is zero.

The maximum amount of entanglement that can be extracted between one mode in $A$ and the mode $B$, however, is certainly not zero. In this particular example, it is clear that the mode in $A$ most entangled with $B$ is the mode $A_2$, since $A_2B$ are assumed to form a partner-mode pair. What is happening here is that the partner mode of the most mixed normal mode of subregion $A$ happened to be fully concentrated outside the support of the subregion $B$, and thus all the entanglement between subregions $A$ and $B$ turned out to consist of entanglement between $B$ and the \emph{second} most mixed normal mode of $A$.

This toy example points to the fact that the optimal choice of modes for entanglement extraction between arbitrary regions will depend heavily on the spatial distribution of partner-mode pairs. Even though the most mixed normal mode of a subregion and its complement may be the best choices for entanglement harvesting if one has access to the entirety of space, other modes may be crucial when considering entanglement between regions separated by some finite distance. An important next step is to study in more detail how to make use of the spatial localization of partner-mode pairs, and more importantly, learn whether it is possible to formulate a general principle for the optimal choice of modes for entanglement extraction that also applies to mixed Gaussian states.

On another note, even though we used the vacuum state of a free scalar field in flat spacetimes for all the examples in the paper, it is possible to extend this approach to curved spacetimes in generally covariant form. In this case, the field's two-point function $W(\mathsf{x}, \mathsf{y}) = \langle \hat{\phi}(\mathsf{x})\hat{\phi}(\mathsf{y})\rangle$ for arbitrary spacetime points $\mathsf{x}$ and $\mathsf{y}$ plays a role analogous to that which the covariance matrix plays in lattice calculations. In globally hyperbolic spacetimes with a global timelike Killing vector, it is natural to construct the two-point function as the time-evolved version of the correlation functions between field and conjugate momentum on a fixed-time hypersurface. Those equal-time correlation functions (which are nothing more than a continuum version of a covariance matrix) can be established by defining the field state as the ground state of the Hamiltonian that generates time evolution with respect to the timelike isometry of the background geometry. However, the superficially covariance-breaking step, where we single out a preferred foliation of spacetime in terms of the timelike coordinate associated to the orbits of the isometry, does not have to be seen as fundamental. One can equivalently interpret the state of the field as being completely defined (under the assumption of Gaussianity) by the knowledge of the two-point function between arbitrary spacetime points, which precludes any privileged choice of foliation of spacetime. This insight has been fruitful in the causal set approach to quantum gravity, by allowing one to make sense of a covariant notion of entanglement entropy for general spacetime regions purely in terms of the field's two-point function~\cite{SorkinEntropy1, SorkinEntropy2, SorkinEntropy3}. By extending the ideas explored in this paper, it should be possible to adapt this strategy to more general entanglement measures between disconnected spacetime regions in curved spacetimes as well. 

Finally, while this paper is using particle detectors as a mathematical tool to simplify and make the calculation of field entanglement computationally accessible even in complicated setups, one should not neglect that this approach directly connects with the way we model the interaction of realistic particle detectors (such as atomic systems) with realistic quantum fields (such as the electromagnetic field). These methods can therefore in principle be naturally implementable in experiments, or even simulated in quantum computers. There is, however, one important difference between the form of field-detector coupling that is usually employed for entanglement harvesting, and the one we are proposing in this paper. In typical setups in relativistic quantum information, the coupling between a detector and the field of interest is derived from physical constraints on the detector's internal structure, as well as on an underlying principle that dictates the form of the interaction. In the paradigmatic example of the light-matter interaction, for instance, the energy gap and smearing function of the detector (typically a hydrogenlike atom) are set by the electron's internal energy levels and wavefunctions, and the way the atom interacts with an external electromagnetic field is fundamentally derived from minimal coupling. In contrast, the kind of coupling prescribed by an interaction that leads to time evolution of the form~\eqref{gaussianswap} is in general quite different. However, it is in principle possible to design an effective coupling by considering arrays of particle detectors, spatially distributed as to mimic the correct smearing function of the normal mode of interest, and engineer a swap between a specific localized mode of the field and one mode of the multiple detector system. Similar protocols have already been proposed in the context of quantum energy teleportation~\cite{Funai2017}, and while the focus of this paper is not to suggest experimental implementations of the measurement of the entanglement structure of a QFT, this is an interesting avenue that warrants future exploration.

In summary, the present work provides a systematic study of what kind of interaction one should aim for in order to faithfully capture the entanglement originally present in a quantum field. The conceptual framework proposed here may provide a useful organizing principle for the understanding of entanglement in quantum field theory, as well as to how it can be exchanged with localized probes.



\section{Acknowledgements}
The authors thank Ivan Agull\'{o}, Patricia Ribes Metidieri and Sergi Nadal for our very enlightening exchanges. We also would like to thank Robert H. Jonsson for helpful discussions. EMM acknowledges the support of the NSERC Discovery program as well as his Ontario Early Researcher Award. BSLT and JPG acknowledge support from the Mike and Ophelia Lazaridis Fellowship. JPG also received the support of a fellowship from “La Caixa” Foundation (ID 100010434, with fellowship code LCF/BQ/AA20/11820043). Research at Perimeter Institute is supported in part by the Government of Canada through the Department of Innovation, Science and Economic Development Canada and by the Province of Ontario through the Ministry of Colleges and Universities.

\appendix

\section{Primer on Gaussian quantum mechanics}\label{GQMSection}

The main workhorse throughout this work, which is the basis of our analytical and numerical treatment of the problem, is an approach to the description of bosonic quantum systems known as \emph{Gaussian formalism} or \emph{Gaussian quantum mechanics}. This section will contain a broad overview of the main technical concepts and tools from Gaussian Quantum Mechanics that we employ in this work; for a more detailed introduction, see, e.g.,~\cite{EduardoGQM} and references therein.

A general quantum system consisting of $n$ bosonic degrees of freedom is fully described by a set of $n$ positions $\hat{Q}^i$ and $n$ momenta $\hat{P}_i$, which are operators (collectively referred to as \emph{quadratures}) satisfying the canonical commutation relations,
\begin{align}\label{commrelations}
    \comm{\hat{Q}^i}{\hat{Q}^j} &= 0, \nonumber \\
    \comm{\hat{P}_i}{\hat{P}_j} &= 0, \nonumber \\
    \comm{\hat{Q}^i}{\hat{P}_j} &= \ii \delta^{i}_{j}\mathds{1}.
\end{align}
We will refer to any canonically conjugate pair $(\hat{Q}^i, \hat{P}_i)$ (\emph{no} summation in $i$ implied!) as a \emph{mode}. The commutation relations~\eqref{commrelations} can be recast in a more concise form as
 \begin{equation}\label{covcommrelations}
        \comm{\hat{\Xi}^\alpha}{\hat{\Xi}^\beta} = \ii \Tilde{\Omega}^{\alpha\beta}\mathds{1},
    \end{equation}
    where we defined the operator-valued vector
    \begin{equation}
    \hat{\bm{\Xi}} = (\hat{Q}^1, \hat{P}_1, \dots, \hat{Q}^n, \hat{P}_n)^\intercal
    \end{equation}
    and the matrix
    \begin{equation}\label{inversesympform}
    (\tilde{\Omega}^{\alpha\beta}) = \bigoplus_{i = 1}^n \begin{pmatrix}0 & 1 \\ -1 & 0 \end{pmatrix}.
    \end{equation}
    One can fully represent any quantum state $\hat{\rho}$ of a system with $n$ bosonic degrees of freedom by means of the function
    \begin{equation}\label{WignerFunctionFinalPhaseSpace}
    W_{\hat{\rho}}(\bm{\xi}) = \dfrac{1}{(2\pi )^{2n}}\int \dd^{2n} \xi'\,\,e^{\ii\, \xi^\alpha\Tilde{\Omega}_{\alpha\beta}\xi'^\beta}\Tr\left(\hat{\rho}\,e^{\ii\,\xi'^\alpha\Tilde{\Omega}_{\alpha\beta}\hat{\Xi}^\beta}\right),
    \end{equation}
    where $\bm{\xi}$ is now a \emph{real-valued} phase space vector,
    \begin{equation}
        \bm{\xi} = (q^1, p_1, \dots, q^n, p_n)^\intercal,
    \end{equation}
    the matrix $\Tilde{\Omega}_{\alpha\beta}$---known as the \emph{symplectic matrix}---is the inverse of Eq.~\eqref{inversesympform},
    \begin{equation}\label{sympform}
    (\tilde{\Omega}_{\alpha\beta}) = \bigoplus_{i = 1}^n \begin{pmatrix}0 & -1 \\ 1 & 0 \end{pmatrix}
    \end{equation}
    and $\dd^{2n} \xi' = \dd^n q' \dd^n p'$ is the canonical phase-space volume element, with the integral in Eq.~\eqref{WignerFunctionFinalPhaseSpace} thus representing a full phase-space integral. The object $W_{\hat{\rho}}(\bm{\xi})$ is known as the \emph{Wigner function}. 
    
    The Wigner function $W_{\hat{\rho}}(\bm{\xi})$ provides a complete description of the state $\hat{\rho}$ in terms of a real-valued function on phase space. The expectation value on state $\hat{\rho}$ of any operator $\hat{\mathcal{O}}$ corresponding to the Weyl-quantized~\cite{Weyl, Weyl2} version of a phase-space function $O(\bm{\xi})$ can be readily computed as
    \begin{equation}
        \expval{\hat{\mathcal{O}}}_{\hat{\rho}}\equiv \Tr\left(\hat{\rho}\,\hat{\mathcal{O}}\right) = \int \dd^{2n}\xi\, W_{\hat{\rho}}(\bm{\xi})O(\bm{\xi}).
    \end{equation}
    Moreover, if we split the phase space vector $\bm{\xi}$ explicitly into positions $\bm{q}$ and momenta $\bm{p}$, the marginals of $W_{\hat{\rho}}(\bm{\xi})$ with respect to $\bm{p}$ precisely give the probability distribution for $\bm{q}$ on state $\hat{\rho}$, and vice-versa. For a completely general state $\hat{\rho}$, however, there may be regions of phase space where $W_{\hat{\rho}}(\bm{\xi})$ takes negative values; therefore, in general the Wigner function cannot be seen as a bonna-fide probability distribution on phase space---which is why it is often referred to as a \emph{quasi}-probability distribution instead.
    
    A very important class of states consists of those whose Wigner functions are \emph{Gaussian}, for which there exist a phase space vector $\bm{\xi}_0$ and a $2n \times 2n$ symmetric matrix $\sigma$ such that
    \begin{equation}\label{gaussianstates}
    W_{\hat{\rho}}(\bm{\xi}) = \dfrac{1}{\pi^n\sqrt{\det(\sigma)}} \exp\left[-\left(\bm{\xi} - \bm{\xi}_0\right)^\intercal\sigma^{-1}\left(\bm{\xi} - \bm{\xi}_0\right)\right].
    \end{equation}
    Since the quadratures $\hat{\Xi}^\mu$ correspond directly to the Weyl-quantized version of the phase space variables $\xi^\mu$, it is straightforward to check that
    \begin{align}
    \xi_0^\mu =& \expval{\hat{\Xi}^\mu}_{\hat{\rho}}, \\
    \sigma^{\mu\nu} =& \expval{\hat{\Xi}^{\mu}\hat{\Xi}^{\nu} + \hat{\Xi}^{\nu}\hat{\Xi}^{\mu}}_{\hat{\rho}} - 2\xi_0^\mu \xi_0^\nu\nonumber \\
     =& \expval{\left(\hat{\Xi}^{\mu} - \xi_0^\mu\right)\left(\hat{\Xi}^{\nu} - \xi_0^\nu\right)}_{\hat{\rho}} \nonumber \\ &+ \expval{\left(\hat{\Xi}^{\nu} - \xi_0^\nu\right)\left(\hat{\Xi}^{\mu} - \xi_0^\mu\right)}_{\hat{\rho}}.\label{covmatrix}
    \end{align}
    The vector $\bm{\xi}_0$ is called the \emph{vector of means}, and $\sigma$ is known as the \emph{covariance matrix}. By usual tricks on Gaussian integration, all higher-order moments of the quadratures can be expressed entirely in terms of the first and second moments $\bm{\xi}_0$ and $\sigma$.
    
    States whose Wigner function takes the form in Eq.~\eqref{gaussianstates} are called \emph{Gaussian states}. They describe a plethora of states found in common systems in quantum optics and quantum field theory, and can be seen as composed of coherent, squeezed, and thermal states of systems with quadratic Hamiltonians. In particular, the vacuum state of a free scalar field, as well as its reduced state on any localized region, is a Gaussian state.
    
    Because Gaussian Wigner functions are always manifestly positive-definite, Gaussian states are examples of quantum states that \emph{can} be understood as probability distributions on phase space, and are therefore in some sense ``classical'' (even though they can still display distinctly quantum features such as entanglement, as we will discuss shortly). In fact, if the state $\hat{\rho}$ is pure, it can be shown that $W_{\hat{\rho}}(\bm{\xi})$ is a  positive-definite function if \emph{and only if} it is Gaussian.\footnote{This fact is known in the literature as \emph{Hudson's theorem}~\cite{hudson, hudson2}.} For a Gaussian state, the positivity requirement---namely, that $\hat{\rho}$ is a positive-semidefinite operator on the Hilbert space---reduces to a condition on the covariance matrix,
    \begin{equation}
        \sigma + \ii\Tilde{\Omega} \geq 0.
    \end{equation}
    This can be seen as a manifestation of the uncertainty principle, which prevents the covariances from all being too small simultaneously, and acts like a ``minimum area'' constraint for the phase space probability distribution described by $W_{\hat{\rho}}(\bm{\xi})$.
    
    By construction, Gaussian states are fully fixed once the vector of means and the covariance matrix of its quadratures are known. Therefore, if a given system is initialized in a Gaussian state and the time evolution preserves Gaussianity, the dynamics becomes effectively finite-dimensional, since one only has to keep track of the time evolution of the vector of means and the covariance matrix. These two main facts yield vast simplifications in the characterization of Gaussian states, both analytically and numerically. What we mean by \emph{Gaussian formalism} or \emph{Gaussian Quantum Mechanics} is then simply the collection of the analytical techniques and results that apply to the study of Gaussian states thanks to these simplifications.
    
    Unitary time evolution that preserves Gaussianity is generated by Hamiltonians that are at most quadratic in the quadrature operators. For a general quadratic Hamiltonian of the form
    \begin{equation}
        \hat{H} = \dfrac{1}{2}\hat{\Xi}^\mu F_{\mu\nu}\hat{\Xi}^\nu + \alpha_{\mu}\hat{\Xi}^\mu,
    \end{equation}
    where $\bm{F}$ is a $2n \times 2n$ Hermitian matrix, and $\bm{\alpha}$ is a real-valued vector, the time evolution operator \mbox{$\hat{U}(t) = \exp(-\ii\hat{H}t)$} is represented in phase space as a linear-affine symplectic transformation on the quadratures,
    \begin{equation}\label{evolutionquadratures}
    \hat{\bm{\Xi}} \mapsto \hat{\bm{\Xi}}(t) = \hat{U}^\dagger(t)\hat{\bm{\Xi}}\hat{U}(t) = \bm{S}\hat{\bm{\Xi}} + \bm{d}\mathds{1},
\end{equation}
    where we have
    \begin{align}
    \bm{S}(t) = \exp\left(\Tilde{\Omega}^{-1}\bm{F}t\right), \\
    \bm{d}(t) = \dfrac{\bm{S}(t) - \mathds{1}}{\Tilde{\Omega}^{-1}\bm{F}}\Tilde{\Omega}^{-1}\bm{\alpha}.
\end{align}
    For any symmetric matrix $F$, one can show that $S(t)$ satisfies
    \begin{equation}\label{symplecticcondition}
        S(t)\Tilde{\Omega}S^\intercal(t) = \Tilde{\Omega}.
    \end{equation}
    The condition~\eqref{symplecticcondition} guarantees that the time evolution preserves the canonical commutation relations~\eqref{covcommrelations}. Any matrix that satisfies~\eqref{symplecticcondition} is called \emph{symplectic}. Under time evolution of the form in Eq.~\eqref{evolutionquadratures}, the vector of means and covariance matrix evolve as
    \begin{align}
    \bm{\xi}_0 \mapsto& \bm{S}(t)\bm{\xi}_0 + \bm{d}(t), \\
    \sigma \mapsto& \bm{S}(t)\sigma \bm{S}^\intercal(t).
\end{align}

The mixedness and entanglement contained in a Gaussian state are fully determined by the covariance matrix $\sigma$. Since one can always set the vector of means $\bm{\xi}_0$ to be zero by a unitary phase-space displacement that acts locally on each mode and therefore does not affect the entanglement across modes, we will from now on take $\bm{\xi}_0 = 0$. 

Thanks to Williamson's theorem~\cite{Williamson}, for any covariance matrix $\sigma$, there always exists a symplectic transformation $S$ such that $\sigma$ can be written as
\begin{equation}
    \sigma = S\sigma_D S^\intercal
\end{equation}
where $\sigma_D$ is of the form
\begin{equation}
    \sigma_D = \bigoplus_{i = 1}^n \begin{pmatrix}\nu_i & 0 \\ 0 & \nu_i \end{pmatrix}.
\end{equation}
The set $\{\nu_i\}_{i=1}^n$ is called the \emph{symplectic spectrum} of $\sigma$, and the elements of the symplectic spectrum are known as the \emph{symplectic eigenvalues}. Williamson's theorem implies that one can always find a basis of canonical coordinates on phase space (i.e., a choice of modes) such that the covariance matrix in that basis is fully decoupled. We refer to this set of modes as the \emph{normal mode basis}, since they are mathematically analogous to the set of normal modes in terms of which a given time-independent quadratic Hamiltonian becomes just a sum of decoupled single-mode Hamiltonians.

The constraint \mbox{$\sigma + \ii \Tilde{\Omega} \geq 0$}---which, we recall, is the condition for a covariance matrix to describe a positive-semidefinite Gaussian quantum state $\hat{\rho}$---can be very easily rephrased in terms of the symplectic spectrum as the condition that $\nu_i \geq 1$ for every $i$. A pure Gaussian state will have all of its symplectic eigenvalues equal to $1$; if any $\nu_i$ is larger than $1$, then the state is mixed. 

Quantitative measures of mixedness of a Gaussian state can always be expressed in terms of its symplectic spectrum. For instance, the von Neumann entropy $S(\hat{\rho})$ of a Gaussian state with symplectic eigenvalues $\{\nu_i\}$ is given by
\begin{align}
    S(\hat{\rho}) &\equiv -\Tr\left(\hat{\rho}\log\hat{\rho}\right) \nonumber \\
    &= \sum_{i}\left[\dfrac{\nu_i + 1}{2}\log\left(\dfrac{\nu_i+1}{2}\right)-\dfrac{\nu_i - 1}{2}\log\left(\dfrac{\nu_i-1}{2}\right)\right].
\end{align}

Consider now a partition of a Gaussian state into two subsystems $A$ and $B$ with $N_A$ and $N_B$ modes respectively, and denote the covariance matrix of the joint system by $\sigma_{AB}$. Then, the reduced states of subsystems $A$ and $B$ will clearly also be Gaussian, with their respective covariant matrices being obtained from $\sigma_{AB}$ in a very simple manner. If we order the basis of our coordinate system such that the first $2N_A$ components of the phase space vector $\bm{\xi}$ refer to the degrees of freedom in $A$ and the next $2N_B$ components are relative to subsystem $B$, then the covariance matrix $\sigma_A$ for the reduced state of $A$ is just the $2N_A\times 2N_A$ upper left block of $\sigma_{AB}$; similarly, the covariance matrix $\sigma_B$ for the reduced state of $B$ will just contain the $2N_B\times 2N_B$ bottom right block of $\sigma_{AB}$.

If the overall state of the joint system $\sigma_{AB}$ is pure, then the entanglement between $A$ and $B$ can be fully characterized by the mixedness of each subsystem. The entanglement in this case is quantified by the entanglement entropy, which is given by the von Neumann entropy of either $A$ or $B$. If the joint state of $AB$ is mixed, however, the mixedness of either subsystem is no longer a good diagnostic for entanglement. For that purpose, the measure of entanglement we will adopt is the \emph{logarithmic negativity}~\cite{Vidal2001,Plenio2005}, defined for any bipartite quantum state $\hat{\rho}_{AB}$ as
\begin{equation}
    E_{\mathcal{N}}(\hat{\rho}) = \log\bignorm{\hat{\rho}^{T_B}}_1
\end{equation}
where $\bignorm{\hat{\mathcal{O}}}_1 \coloneqq \Tr\sqrt{\hat{\mathcal{O}}^\dagger\hat{\mathcal{O}}}$ is the trace norm of the operator $\hat{\mathcal{O}}$, and $\hat{\rho}^{T_B}$ is the result of taking the partial transpose of the original state $\hat{\rho}_{AB}$ with respect to subsystem $B$. $ E_{\mathcal{N}}(\hat{\rho})$ is an entanglement monotone---i.e., a quantity that vanishes for separable states, does not change under local unitaries, and does not increase under local operations and classical communication---which provides an upper bound on distillable entanglement and is far easier to compute in practice than all the other standard measures of entanglement for mixed states.

On phase space, transposition amounts to time reversal, which acts on the quadrature operators by reversing the sign of $\hat{P}_i$ and leaving $\hat{Q}^i$ unchanged. The transformation to the covariance matrix resulting from partial transposition with respect to subsystem $B$ is therefore
\begin{equation}
    \tilde{\sigma}_{AB} = \left(\mathds{1}_{2N_A}\oplus T_B\right)\sigma_{AB}\left(\mathds{1}_{2N_A}\oplus T_B\right)
\end{equation}
where $\mathds{1}_{2N_A}$ is the $2N_A\times 2N_A$ identity matrix, and $T_B$ is given by
\begin{equation}
    T_B = \bigoplus_{i = 1}^{N_B} \begin{pmatrix}1 & 0 \\ 0 & -1 \end{pmatrix}.
\end{equation}
For a bipartite Gaussian state, the logarithmic negativity between subsystems $A$ and $B$ is fully determined by the symplectic spectrum of $\tilde{\sigma}_{AB}$. The statement that $\hat{\rho}^{T_B}$ is not positive semidefinite then turns into the statement that some of the symplectic eigenvalues of $\tilde{\sigma}_{AB}$ are smaller than $1$. In this case, the logarithmic negativity is given by
\begin{equation}\label{lognegativitygaussian}
    E_{\mathcal{N}}(\hat{\rho}) = \sum_{i}F(\tilde{\nu}_i),
\end{equation}
where $\{\tilde{\nu}_i\}$ is the symplectic spectrum of $\tilde{\sigma}_{AB}$, and we define $F(x) = -\log(x)$ for $x\in(0,1]$ and $F(x)=0$ for $x>1$. 
\section{Proof of Equation~\eqref{CitationNeeded} and interlacing theorem for symplectic eigenvalues}\label{AppendixProof}

The symplectic eigenvalues of the partial transpose of a bipartite system $AB$ whose covariance matrix for the partially transposed state is $\Tilde{\sigma}_{AB}$ have a particularly simple expression when the full state $\rho_{AB}$ is pure. Let us first consider that $A$ and $B$ are made of one single mode. Up to local symplectic transformations, the covariance matrix of subsystem $A$ can be expressed as
\begin{equation}
    \sigma_{A} = \begin{pmatrix} \nu & 0 \\ 0 & \nu
    \end{pmatrix}.
\end{equation}
One purification of this covariance matrix is given by 
\begin{equation}\label{twomodesqueezedcovmatrix}
    \sigma_{AP} = \begin{pmatrix} \nu & 0 & \sqrt{\nu^2 - 1} & 0 \\
    0 & \nu & 0 & -\sqrt{\nu^2-1} \\
    \sqrt{\nu^2-1} & 0 & \nu & 0 \\
    0 & -\sqrt{\nu^2 - 1} & 0 & \nu
    \end{pmatrix},
\end{equation}
and any other purification can be obtained from $\sigma_{AP}$ by a local symplectic transformation acting solely on the purifying system $P$. Since local unitaries on $P$ do not alter the entanglement between $A$ and its purifying system, we can use the covariance matrix $\sigma_{AP}$ to compute the entanglement between $A$ and $B$. By taking the partial transpose (i.e., by performing a local time reversal) on $P$, we obtain the covariance matrix
\begin{equation}\label{twomodesqueezedtranspose}
    \tilde{\sigma}_{AP} = \begin{pmatrix} \nu & 0 & \sqrt{\nu^2 - 1} & 0 \\
    0 & \nu & 0 & \sqrt{\nu^2-1} \\
    \sqrt{\nu^2-1} & 0 & \nu & 0 \\
    0 & \sqrt{\nu^2 - 1} & 0 & \nu
    \end{pmatrix}.
\end{equation}
The smallest symplectic eigenvalue of $\tilde{\sigma}_{AP}$ (which is the only one that can be smaller than $1$, and is therefore the only one that can contribute to the logarithmic negativity) is then found to be
\begin{equation}
    \tilde{\nu}_- = \nu - \sqrt{\nu^2-1}.
\end{equation}
We see that any $\nu>1$ will lead to $\tilde{\nu}_-<1$. Moreover, higher values of $\nu$ correspond to lower values of $\tilde{\nu}_-$, and therefore higher values of the logarithmic negativity. This is just a manifestation of the fact that, for an overall pure state, more mixedness in the reduced state on one subsystem signals more entanglement across the bipartition.

This argument generalizes quite naturally to bipartitions with arbitrary numbers of modes. Consider now that $A$ and $B$ have $N_A$ and $N_B$ modes respectively, where we can assume $N_A\leq N_B$ without loss of generality. Due to Williamson's theorem, there is a basis of normal modes for region $A$ such that the covariance matrix takes the form
\begin{equation}
    \sigma_A = \bigoplus_{i=1}^{N_A}\begin{pmatrix}\nu_i & 0 \\ 0 & \nu_i
    \end{pmatrix}.
\end{equation}
Once again, knowing that the overall covariance matrix $\sigma_{AB}$ is pure and using the purification~\eqref{twomodesqueezedcovmatrix} mode by mode, we find that one can pick a basis of modes in the complement of $A$ such that the full covariance matrix of $AB$ can be expressed as
\begin{equation}
    \sigma_{AB} = \sigma_{AP}\oplus \mathds{1}_{2\Delta}
\end{equation}
where we have
\begin{equation}
    \sigma_{AP} = \bigoplus_{i=1}^{N_A}\begin{pmatrix} \nu_i & 0 & \sqrt{\nu_i^2 - 1} & 0 \\
    0 & \nu_i & 0 & -\sqrt{\nu_i^2-1} \\
    \sqrt{\nu_i^2-1} & 0 & \nu_i & 0 \\
    0 & -\sqrt{\nu_i^2 - 1} & 0 & \nu_i
    \end{pmatrix}
\end{equation}
and $\Delta = N_B - N_A$. In general, any pair of modes that purify each other, such that the two modes together are in a product state with the rest of the system, comprise what is known in the literature as a partner mode pair~\cite{HottaPartnerMode1, HottaPartnerMode2, HottaPartnerMode3}. We can thus see the covariance matrix $\sigma_{AP}$ as the covariance matrix of the system expressed in the basis of normal modes of $A$ and their respective partners. Also note that if we now trace out the subsystem $A$, the covariance matrix of the remaining subsystem $B$ becomes
\begin{equation}
    \sigma_{B} = \left[\bigoplus_{i=1}^{N_A}\begin{pmatrix}\nu_{i} & 0 \\ 0 & \nu_i\end{pmatrix}\right]\oplus \mathds{1}_{2\Delta}. 
\end{equation}
The passage above thus shows that the partner mode to the $j$-th most mixed normal mode of $A$ is given by the $j$-th most mixed normal mode of $B$, whenever $A$ and $B$ are complementary subsystems in an overall pure Gaussian state.

Now taking the partial transpose over $B$ gives us the covariance matrix
\begin{equation}
    \tilde{\sigma}_{AB} = \tilde{\sigma}_{AP}\oplus \mathds{1}_{2\Delta}
\end{equation}
where $\tilde{\sigma}_{AP}$ can be written as a direct sum partially-transposed two-mode covariance matrices of the form~\eqref{twomodesqueezedtranspose}. This allows us to directly infer that the symplectic eigenvalues of $\tilde{\sigma}_{AB}$ that are smaller than $1$ are given by
\begin{equation}
    \tilde{\nu}_i = \nu_i - \sqrt{\nu_i^2-1}.
\end{equation}
Because the right-hand side is a decreasing function of $\nu_i$, if the symplectic eigenvalues $\nu_i$ of $\sigma_A$ are ordered from largest to smallest, then the symplectic eigenvalues $\tilde{\nu}_i$ of $\tilde{\sigma}_{AB}$ as given above will be ordered from smallest to largest, with the smallest ones contributing more to the logarithmic negativity as given by Eq.~\eqref{lognegativitygaussian}. 

Finally, it is important to mention one basic property of the symplectic spectrum of submatrices of a given covariance matrix obtained by throwing away part of its degrees of freedom. This will come in handy when going through the proof of one of our claims in Section~\ref{OptimalCouplingSection}. The result is the following:

\emph{Interlacing theorem for symplectic eigenvalues}~\cite{interlacingtheorem1, interlacingtheorem}. Let $M_1$ be a $2N \times 2N$ positive definite symmetric matrix, which we can see as a covariance matrix for a system with $N$ modes. Consider the matrix $M_2$ obtained from $M_1$ by deleting the rows and columns corresponding to any given mode of $M_1$. Let $\{\nu_j(M_1)\}_{j=1}^N$ and $\{\nu_j(M_2)\}_{j=1}^{N-1}$ be the list of symplectic eigenvalues of $M_1$ and $M_2$ respectively, both ordered from smallest to largest. It then follows that
\begin{equation}
    \nu_j(M_1)\leq \nu_j(M_2)\leq \nu_{j+2}(M_1), \,\,\, 1\leq j\leq N-2
\end{equation}
as well as
\begin{equation}
    \nu_{N-1}(M_1)\leq\nu_{N-1}(M_2).
\end{equation}
One immediate consequence of the interlacing theorem for symplectic eigenvalues is that the mixedness of any single mode in a given Gaussian state cannot be made lower than that of the least mixed normal mode of the region where the mode is supported. In Section~\ref{OptimalCouplingSection} we show that a similar application of the theorem can also bound the entanglement between any set of modes within a region and its complement, again in terms of the entanglement with the region's normal modes.


\bibliography{references}

\end{document}